\documentclass[twocolumn,superscriptaddress,aps,10pt]{revtex4}
\usepackage{amsfonts}
\usepackage{amsmath}
\usepackage{amssymb}
\usepackage{graphicx}
\usepackage{dcolumn}
\usepackage{times}

\DeclareMathOperator{\sech}{sech}

\begin{document}

\title{
Solitons and their ghosts in $\mathcal{PT}$-symmetric systems with defocusing
nonlinearities
}
\author{V.~Achilleos}
\affiliation{Department of Physics, University of Athens, Panepistimiopolis,
Zografos, Athens 157 84, Greece}
\author{P.G.~Kevrekidis}
\affiliation{Department of Mathematics and Statistics, University of Massachusetts,
Amherst, Massachusetts 01003-4515, USA}
\author{D.J.~Frantzeskakis}
\affiliation{Department of Physics, University of Athens, Panepistimiopolis,
Zografos, Athens 157 84, Greece}
\author{R.\ Carretero-Gonz\'alez}
\affiliation{Nonlinear Dynamical Systems Group,
Computational Science Research Center, and Department of Mathematics and
Statistics, San Diego State University, San Diego, CA 92182-7720, USA}

\begin{abstract}


We examine a prototypical
nonlinear Schr{\"o}dinger model bearing a defocusing
nonlinearity and Parity-Time ($\mathcal{PT}$) symmetry.
For such a model, the solutions
can be identified numerically and characterized in the
perturbative limit of small gain/loss. There we find two fundamental
phenomena. First, the dark solitons
that persist in the presence
of the $\mathcal{PT}$-symmetric potential are destabilized via a
symmetry breaking (pitchfork) bifurcation. Second, the ground state and
the dark soliton die hand-in-hand in a saddle-center bifurcation
(a nonlinear analogue of the  $\mathcal{PT}$-phase transition)
at a second critical value of the gain/loss parameter. The daughter
states arising from the pitchfork are identified as ``ghost states'',
which are not exact solutions of the original system, yet
they play a critical role in the system's dynamics.
A similar phenomenology is also pairwise identified for higher excited
states, with e.g.~the two-soliton structure bearing similar characteristics
to the zero-soliton one, and the three-soliton state having the same
pitchfork destabilization mechanism and saddle-center collision (in this
case with the two-soliton) as the one-dark soliton.
All of the above notions are generalized in two-dimensional settings for
vortices, where the topological charge enforces the destabilization of a two-vortex
state and the collision of a no-vortex state with a two-vortex one,
of a one-vortex state with a three-vortex one, and so on.
The dynamical manifestation of the instabilities mentioned above
is examined through direct numerical simulations.

\end{abstract}

\maketitle

\section{Introduction.}


Over the past decade, and since they were originally proposed
by Bender and co-workers~\cite{bend1,bend2},
systems characterized by $\mathcal{PT}$-symmetric Hamiltonians
have become a subject of intense research efforts.
The interest on
these systems arises from their fundamental property to exhibit real spectra, while non-Hermitian,
thus
providing an intriguing alternative to standard Hermitian quantum mechanics.
In the case of a standard Schr{\"o}dinger-type Hamiltonian, with a generally complex
potential $U$, the $\mathcal{PT}$ symmetry dictates that the potential satisfies
the condition $U(x)=U^{\ast}(-x)$, where $(\cdot)^{\ast}$ stands for
complex conjugation.

An important development in the study of such models was the work
%
of Christodoulides and co-workers~\cite{christo1,christo1b}, who proposed and studied
(both theoretically and experimentally)
{\it optical systems} featuring the $\mathcal{PT}$ symmetry.
In the optics context, however, another key element that comes into play in the physics of such systems
is nonlinearity.
Therefore, the considerations of Refs.~\cite{christo1,christo1b} extended from bright and gap solitons
to linear (Floquet-Bloch) eigenmodes in periodic potentials, examining how these
coherent structures are affected by the genuinely complex, yet
$\mathcal{PT}$-symmetric potentials. More recently, experimental results were reported
both in nonlinear optical systems \cite{salamo,kip}
and electronic analogs thereof~\cite{tsampikos_recent}. These, in turn, have triggered
a wide range of theoretical studies on
nonlinear lattices with either linear ~\cite{kot1,sukh1,kot2,grae1,grae2,kot3,pgk,dmitriev1,dmitriev2,R30add1,R30add2,R30add3,R30add4,R30add5}
or nonlinear~\cite{miron,konorecent,konorecent2} $\mathcal{PT}$-symmetric
potentials and, more recently, on harmonic $\mathcal{PT}$-symmetric potentials \cite{vvk}.


In the above works,  numerous features
extending from bright solitons to defect modes, and from gap solitons
to $\mathcal{PT}$-lattices have been
examined. Nevertheless, the consideration of defocusing nonlinearities,
and especially of dark solitons
has received limited attention; see, e.g., Refs.~\cite{Li,ourshort}.
Another theme that despite its considerable relevance has also
been considered only by a few works is that of higher dimensionality.
For the latter,  the focusing nonlinearity
case~\cite{christo1,christo1b} has been examined, especially so
in the context of lattice potentials.
Our aim herein is to provide a systematic analysis of
$\mathcal{PT}$-symmetric
Hamiltonians exhibiting defocusing
nonlinearities. Particular, and building on the earlier work
of Ref.~\cite{ourshort}, we give a detailed account of
the existence, stability and dynamical properties of the
ground state and first few excited states namely dark solitary waves
and vortex two-dimensional generalizations thereof.
This is done specifically in a prototypical context of the nonlinear
Schr{\"o}dinger type which can, in principle, be relevant both
in the case of Bose-Einstein condensates~\cite{emergent},
as well as in that of nonlinear optics~\cite{agrawal}; see the detailed
justification of the real and imaginary parts of the potential
below.

Our main findings and their presentation is structured as follows.
\begin{itemize}
\item In Sec.~II, we put forth the general model and consider
its physical relevance.
\item In Secs.~III-IV, we turn to generic potentials and the features of
their ground and excited states in one-dimension. We find that odd
excited states (1-soliton, 3-soliton, 5-soliton, etc.) become
subject to a symmetry breaking bifurcation. The pitchfork nature of
this event is rationalized through the introduction of so-called
ghost states which arise from it and which are exact solutions of
the steady state but remarkably not ones such of the original
full problem. Nevertheless, direct numerical simulations clearly
identify the ensuing symmetry breaking and manifest the dynamical
role of the ghost waveforms. One more important feature identified
is the  nonlinear analogue of the  $\mathcal{PT}$-phase transition.
In particular, between the (unstable, for sufficiently strong
gain/loss) saddles of the odd excited states and the centers of
the even states (ground state, 2-soliton, 4-soliton, etc.), there
is a pairwise collision and disappearance (blue-sky or saddle-center)
event. This is strongly reminiscent of the corresponding
transition of linear eigenstates of the Hamiltonian originally
reported in Ref~\cite{bend1}. Direct numerical simulations are
employed in order to identify the evolution dynamics of the unstable
solitary waves states and also their dynamics past the
$\mathcal{PT}$-phase transition  point in Sec.~V.
The ghost states are then examined separately in their own right in Sec.~VI.
\item In Sec.~VII, we illustrate how each of the above
parts of the picture is generalized in the two-dimensional
variant of the relevant model. There, the charge of
the vortex states imposes topological constraints enforcing that
the ground state may only collide and disappear (in a nonlinear
$\mathcal{PT}$-phase transition) with the two-vortex (of opposite charge,
namely dipole) state. Similarly,
the single vortex and triple (of alternating charge) suffer a saddle-center
bifurcation and so on. Prior to these events, a destabilization of
the two-vortex (and three-vortex etc.) states arises through a pitchfork
event creating ghost vortex states. The latter are illustrated dynamically
as well and the structural analogies
of the one- and two-dimensional
settings are explored both in the statics and in the dynamics. Lastly,
Sec.~VIII presents a brief summary of our conclusions and a number
of potential directions for future study.
\end{itemize}.

\section{The model and some analytical insights}
%
Our model, which can be equally applied to a variety of
one-dimensional and even higher dimensional systems
is a nonlinear Schr\"{o}dinger (NLS) equation incorporating a complex potential. This equation,
which finds applications in the contexts of nonlinear optics~\cite{agrawal}
and in the physics of
atomic Bose-Einstein condensates \cite{emergent}, is expressed in the following dimensionless form:
\begin{eqnarray}
i \partial_t u=-\frac{1}{2} \partial_x^2 u + |u|^2 u + [V(x) + i W(x)] u,
\label{PT1}
\end{eqnarray}
where $u$ is a complex field, denoting the electric field envelope in the context of optics
(or the macroscopic wavefunction in BECs),
$t$ denotes the propagation distance (or time in BECs), $x$ is the transverse direction, while
$V(x)$ and $W(x)$ denote, respectively, the real and imaginary parts of the external potential.
For a $\mathcal{PT}$-symmetric Hamiltonian, $V(x)$ and $W(x)$ must be, respectively, an even
and an odd function of $x$, namely:
\begin{equation}
\left\{
\begin{array}{rcl}
V(x)&=&V(-x), \\[1.0ex]
W(x)&=&-W(-x).\\
\end{array}
\right.
\end{equation}
Physically speaking, in the context of optics, $V(x)$ and $W(x)$ represent, respectively,
the spatial profiles of the real and imaginary parts of the refractive index. The above
requirements for the parities of $V(x)$ and $W(x)$ can be met in a case where $V(x)$ has,
e.g., a parabolic profile and $W(x)$ has an anti-symmetric profile
(amounting to equal and opposite gain and loss), as in the experiment of
Ref.~\cite{kip}. In the context of BECs, $V(x)$ represents the external trap (necessary to confine the
atoms \cite{emergent}) and $W(x)$ accounts for a mechanism for injecting and
removing particles in equal rates. The
requirement for the parity of the trap $V(x)$ can easily be met in the case of
e.g., the usual parabolic potential (representing a magnetic trap)
or a double-well potential (representing a combination of a magnetic
trap with a suitable optical lattice)~\cite{emergent}. On the other hand,
$W(x)$ is odd if an equal number of atoms is injected and removed from
spatial regions symmetrically located around the trap center. Such a setting was originally proposed
in Ref.~\cite{klaiman} for a BEC confined in a double-well potential.
Furthermore, passive parity-time symmetric analogs of the double
well system (featuring only loss in one well and no gain)
have also been proposed in the context of the so-called
open Bose-Hubbard dimer~\cite{tsampas}; see also Refs.~\cite{grae1,grae2}.
More recently, a very large volume of activity has focused on such
double-well potentials in a balanced gain-loss (i.e.,
$\mathcal{PT}$-symmetric) form; see, for instance,
Refs.~\cite{R34,R44,R45,R46}. In such contexts, some of the notions
presented below, such as the ghost states and their emergence
from suitable bifurcations and dynamical relevance, can be both
numerically manifested, as well as analytically demonstrated.

We seek standing wave solutions of Eq.~(\ref{PT1}) in the form $u=\rho(x) \exp[i \phi(x)-i\mu t]$, where the real
functions $\rho(x)$ and $\phi(x)$, and the real constant $\mu$ represent, respectively, the amplitude, phase and
propagation constant (in optics) or chemical potential (in BECs). Substituting this ansatz into Eq.~(\ref{PT1}),
and separating real and imaginary parts, we obtain the
following coupled boundary-value problems (BVPs):
\begin{eqnarray}
\mu \rho &=& -\frac{1}{2} (\rho_{xx} - \rho \phi_x^2) + \rho^3 + V(x) \rho
\label{PTdark2}
\\
2 W(x) \rho^2 &=& (\rho^2 \phi_x)_x,
\label{PTdark3}
\end{eqnarray}
where subscripts denote partial derivatives. For generic potentials $V(x)$ and $W(x)$, the relevant
$\mathcal{PT}$-symmetric problem involves solving the BVPs for $\rho$ and $\phi$. Notice
the critical role of the imaginary part of the potential in dictating the phase profile
$\phi(x)$.
We will chiefly focus on the
case of a real parabolic potential,
\begin{equation}
V(x)=\frac{1}{2}\Omega^2 x^2,
\label{V}
\end{equation}
with strength $\Omega$,
modeling the transverse distribution of the refractive index
(or the external trap in BECs) as mentioned above, while the imaginary part
$W(x)$ will be considered to be an odd, localized function of space,
of spatial width $\ll \Omega^{-1}$. Such a form of $W(x)$ is consistent with the experimental
work of Ref.~\cite{kip} in a nonlinear optics setup
and could be relevant to an effective description of atom loss/gain mechanisms in trapped BECs.
Our analysis will be general (independent of the particular form of $W(x)$), and we will
showcase our results in the case of the following prototypical example:
\begin{equation}
W(x)=\varepsilon x \exp(-x^2/2),
\label{W}
\end{equation}
where $\varepsilon$ is a parameter setting the magnitude of the imaginary potential
(a generalization of this model in two-dimensions will be studied in Sec.~\ref{sec:2D}).
We should mention that our analytical approximations (see below) can also be applied for other choices,
e.g., when $W(x)$ takes the form $W(x)=\varepsilon\, {\rm sech}^2(x)\tanh(x)$; in such cases, we have
checked that our results remain qualitatively similar to the ones that we will present below.

We conclude this section by noting the following. Below we will study the ground state and excited
states of the system (in the form of dark solitons). In that regard, it is relevant to consider the
evolution of the physically relevant quantity $N=\int|u|^2 dx$, which represents power in
optics or number of atoms in BECs. Employing Eq.~(\ref{PT1}), it is straightforward to find
that $dN/dt$ is governed by the equation:
\begin{equation}
\frac{dN}{dt}=2\int_{-\infty}^{+\infty} |u|^2W(x) dx.
\label{dNdt}
\end{equation}
Thus, since $W(x)$ is odd, it is obvious that the power is conserved
as long as the square modulus profile of the ground state or of
the excited states remains even. Below, we will show that for genuinely
stationary states this is the case,
indeed and we will examine the important consequences
of Eq.~(\ref{dNdt}) on the bifurcations and dynamics of the system at hand.

\section{Ground state and single dark soliton}

\begin{figure}[tbp]
\includegraphics[width=8cm]{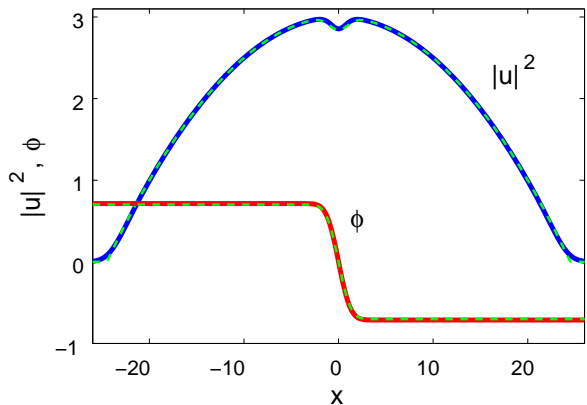}
\caption{(Color online)
The density [solid (blue) line] and phase [solid (red) line] of the numerically obtained
TF background compared to the prediction of Eqs.~(\ref{tf23})
[dashed (green) lines].
The parameters values are: $\mu=3$, $\Omega=0.1$ and $\varepsilon=0.3$.}
\label{fig1}
\end{figure}

\subsection{Ground state}

First, we will study the existence and stability of the most fundamental state of the system, namely the ground state.
The latter is sought as a stationary solution of Eq.~(\ref{PT1}) in the form $u= u_b(x) \exp(-i\mu t)$
(where $\mu$ is the propagation constant or the chemical potential in BECs),
with the background field $u_b$ obeying the equation:
\begin{eqnarray}
-\frac{1}{2}\partial_x^2 u_{b} + |u_b|^2 u_b + [V(x) + i W(x)]u_b-\mu u_b=0.
\label{PT2}
\end{eqnarray}
For a sufficiently small imaginary potential, $W(x)=\varepsilon \tilde{W}(x)$
[with ${\rm max} \{|\tilde{W}(x)|\} = O(1)$], where $\varepsilon \ll 1$, and when
the inverse width $\Omega^{-1}$ of $V(x)$ is sufficiently large so
that $\Omega \sim \varepsilon $,
we may find  an approximate solution of Eq.~(\ref{PT1})
in the Thomas-Fermi (TF) limit \cite{emergent,review}. This solution is of the form:
\begin{eqnarray}
u_b(x)=\left[\sqrt{\mu}+f\left(x\right)\right]\exp[i\phi(x)],
%
\label{TF1}
\end{eqnarray}
where the amplitude and phase $f(x)$ and $\phi(x)$
(considered to be small, of order $\varepsilon^2$ and $\varepsilon$, respectively) are given by:
\begin{eqnarray}
f(x)&=&-\frac{1}{2\sqrt{\mu}} \left(V +2 {\cal W}^2\right),
\\[1.0ex]
\phi(x)&=& 2\int {\cal W}\, dx.
\label{tf23}
\end{eqnarray}
where ${\cal W}=\int W dx$, and we have neglected terms of order $O(\varepsilon^3)$ [notice that the
integral in Eq.~(\ref{tf23}) is an indefinite one]. Contrary
to the conservative case ($\varepsilon=0$) \cite{emergent}, this TF background
is characterized by a density dip located at the center, $x=0$, and
a nontrivial (tanh-shaped) phase distribution. Both features are shown in
Fig.~\ref{fig1}, where
the analytical result (dashed lines) is compared with the numerical one (solid lines);
the agreement between the two is excellent. Note that we have checked that the above solution stays close to the numerically found ground state of the system up to the order $\mathcal{O}(\varepsilon^2)$ \cite{ourshort}.

The evolution of the density of the TF background
indicates
that the ground state is stable (see more details in the analysis below).
Importantly, a linear stability ---Bogoliubov-de Gennes (BdG)--- analysis (see, e.g., Ref.~\cite{review}) justifies
the above argument, showing that the background $u_b(x)$ is
indeed stable against small perturbations.

\begin{figure}[tbp]	
\includegraphics[width=8cm]{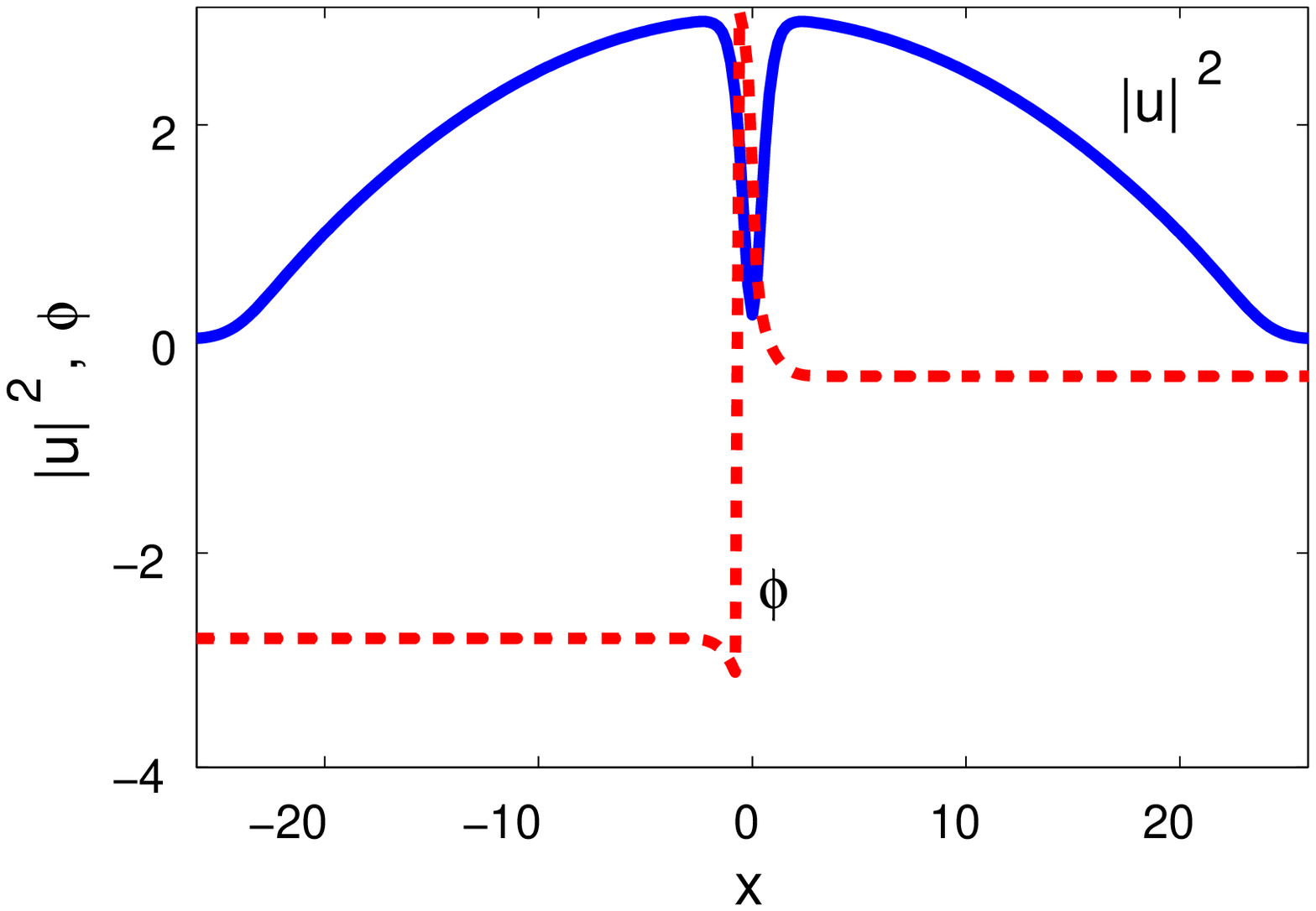}
\includegraphics[width=8.5cm]{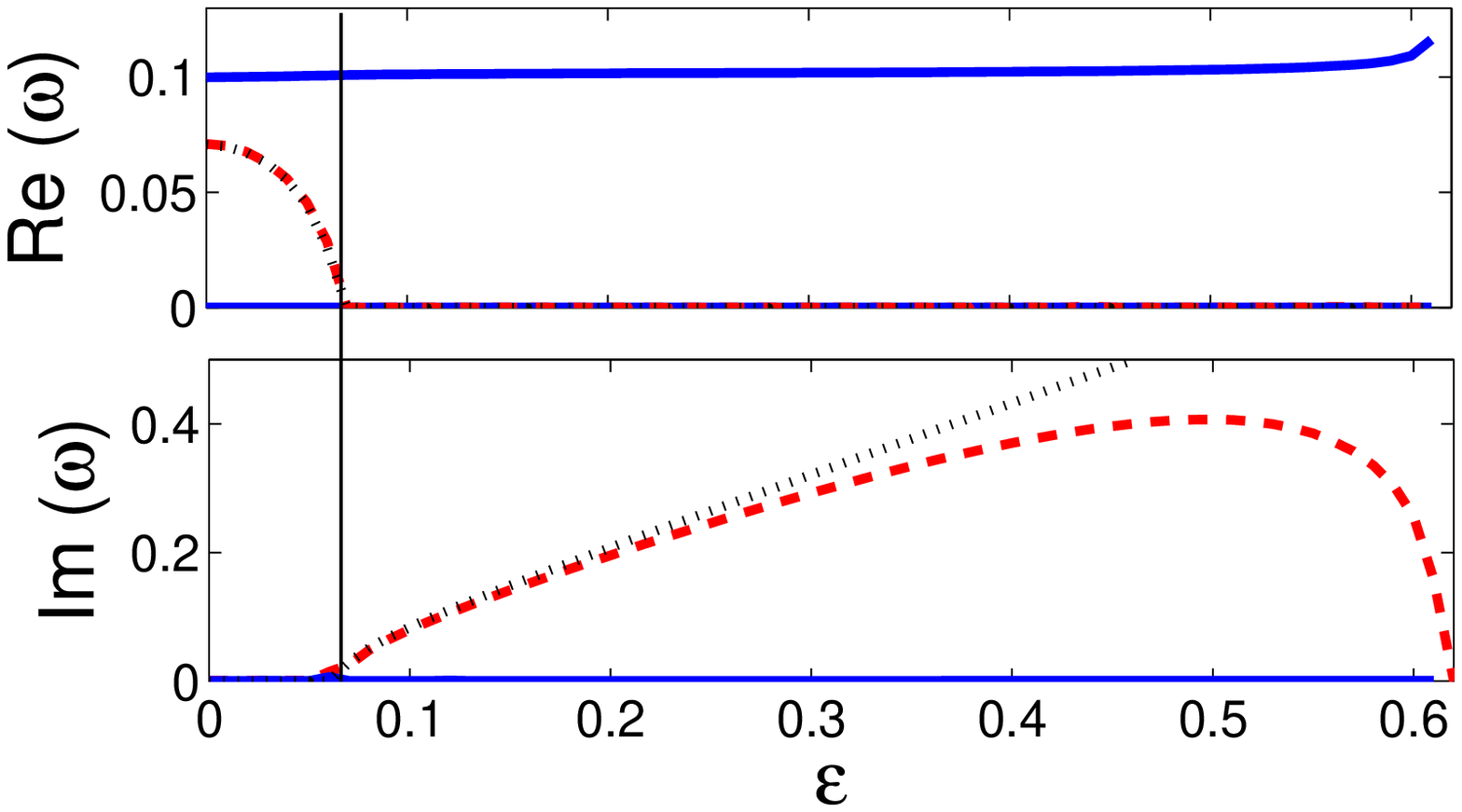}
\caption{(Color online) Top panel: The density [solid (blue) line] and phase
[dashed (red) line] of the first excited (single dark soliton) state for $\varepsilon=0.3$.
Middle and bottom panels: The linear spectrum of the single dark soliton branch: middle (bottom) panel
shows the real (imaginary) part of the lowest eigenfrequencies $\omega$, with respect to the amplitude
$\varepsilon$. The dashed (red) line depicts the mode $\omega_\alpha$, which coincides with the anomalous mode for $\varepsilon=0$; while the dotted (black) line is the analytical result of Eq.~(\ref{solev}). The vertical line shows the point $\varepsilon_{\rm cr}^{(1)}$, where the bifurcation emerges. The parameters used are: $\mu=3$ and $\Omega=0.1$.}
\label{fig2}
\end{figure}

\subsection{Single dark soliton}

Apart from the ground state, excited states of the system ---in the form of stationary
dark solitons--- can also be found numerically, by means of a fixed point algorithm
(Newton's method). A pertinent example of the form of a single dark soliton
 is shown in
the top panel of Fig.~\ref{fig2}.

To analyze the dynamics of such a single dark soliton, described by function $\upsilon(x,t)$, on top of the
TF background, we introduce the product ansatz:
$u=u_b(x)\upsilon(x,t)$ into Eq.~(\ref{PT1}) (i.e., we decompose the
solution into a soliton and its background).
Then, we employ the expressions (\ref{TF1})--(\ref{tf23}) for $u_b$, and
simplify the resulting equation for $\upsilon(x,t)$
by keeping only leading-order terms, up to order $\mathcal{O}(\varepsilon^2)$ [recall that the function $f(x)$ is of order $\mathcal{O}(\varepsilon^2)$]. To this end, using the scale transformations
$t \rightarrow \mu t$ and $x \rightarrow  \sqrt{\mu} x $, we obtain the following
perturbed NLS equation:
\begin{eqnarray}
i\partial_t \upsilon+\frac{1}{2}\partial_{x}^2 \upsilon+\upsilon(1-|\upsilon|^2)=P(\upsilon),
\label{nlsd}
\end{eqnarray}
where the functional perturbation $P(\upsilon)$
is given by
\begin{eqnarray}
P(\upsilon)&=&\mu^{-2} \left[
(1-|\upsilon|^2)\upsilon\left(V+2{\cal W}^2\right)
\phantom{\frac{1}{2}}
\right.
\notag
\\[1.0ex]
&&
\left.
+ \upsilon_x \left(\frac{1}{2} V_x -2(W-i) {\cal W} \right)\right].
\end{eqnarray}
Applying the perturbation theory for dark solitons \cite{review,yuri},
we seek a solution of Eq.~(\ref{nlsd}) in the form of the dark soliton of the
unperturbed system ($P(\upsilon)=0$):
\begin{equation}
\upsilon(x,t)=\cos \varphi(t) \tanh \xi +i \sin \varphi(t).
\label{ds}
\end{equation}
Here, $\xi \equiv \cos \varphi(t) \left[ x-x_0(t)\right]$, while the unknown, slowly-varying functions
$\varphi(t)$ and $x_0(t)$ represent, respectively, the phase ($|\varphi| \le \pi/2$)
and center of the soliton. In the framework of the adiabatic approximation,
the perturbation-induced
evolution equations for $x_0(t)$ and $\varphi(t)$ read:
\begin{eqnarray}
\frac{dx_0}{dt}&=&\sin \varphi(t) \label{dxdt} \\
\frac{d\varphi}{dt}&=&-\frac{1}{2}\partial_x V - \int\sech^4(\xi)\left[\tanh(\xi){\cal W}^2
+W {\cal W} \right] dx, \nonumber \\
\label{integ}
\end{eqnarray}
where we have assumed almost black solitons with sufficiently small phase angles.
Thus, generally, for a given imaginary potential $W(x)$, and by calculating the above integral, we can
derive an equation of motion for the dark soliton center $x_0$ in the form:
\begin{eqnarray}
\frac{d^2x_0}{dt^2}=-\frac{\partial V_{{\rm eff}}}{\partial x_0},
\label{pot}
\end{eqnarray}
where $V_{{\rm eff}}$, is the effective potential felt by the single soliton.

In the case where the imaginary part of the potential takes the form of Eq.~(\ref{W}),
we numerically calculate the integral in Eq.~(\ref{integ}) and find that the effective potential can be approximated as:
\begin{eqnarray}
V_{{\rm eff}}(x_0)=\frac{1}{2}\left(\frac{\Omega}{\sqrt{2}}\right)^2 x_0^2
+\frac{\varepsilon^2 c_1}{4c_2}\sech^4(c_2 x_0).
\label{Veff}
\end{eqnarray}
where $c_1 \approx 1.9$ and $c_2 \approx 0.6$. Notice that the effective potential contains essentially two contributions, the first from the external parabolic trap (as is the case in the context of BECs \cite{emergent,review,theocharis}a), and the second from the presence of the imaginary potential $W(x)$.

To examine the stability of the equilibrium at $x_0=0$ (i.e., the stability of a quiescent black soliton
located at the trap center), we Taylor expand the effective potential of Eq.~(\ref{Veff})
around $x_0=0$, and find ---to leading order--- the following equation of motion for the
soliton center \cite{ourshort}:
\begin{eqnarray}
\frac{d^{2}x_{0}}{dt^{2}}&=& -\omega_{\rm osc}^2\, x_0,
\\[1.0ex]
\omega_{\rm osc}^2 &\approx&  \left( \frac{\Omega}{\sqrt{2}}\right)^2-1.14 \varepsilon^2.
\label{solev}
\end{eqnarray}
Equation~(\ref{solev}) implies that if the amplitude $\varepsilon$ of $W(x)$ is less than a critical value
$\varepsilon_{\rm cr}^{(1)}\approx 0.66\,\Omega$, the soliton performs oscillations in the complex potential with frequency
$\omega_{\rm osc}$; on the other hand, if $\varepsilon>\varepsilon_{\rm cr}^{(1)}$ the soliton becomes unstable. In fact, the prediction has the quintessential
characteristics of a pitchfork bifurcation.
For $\varepsilon < \varepsilon_{\rm cr}^{(1)}$, the relevant potential
is monostable at $x_0=0$, while beyond the critical point,
two symmetric minima arise (in this effective potential picture)
and the center position becomes the saddle point that separates
them.

The above prediction about the change of stability of
the critical point at $x_0=0$
has been confirmed numerically, both by means of direct simulations and
by employing a BdG analysis.
The latter reveals that the considered stationary dark soliton is characterized by a mode  $\omega_\alpha$
(initially coinciding with  the anomalous mode~\cite{theocharis,review} for $\varepsilon=0$), which is real for
$\varepsilon<\varepsilon_{\rm cr}^{(1)}$ (in this case, $\omega_\alpha = \omega_{\rm osc}$), and it becomes imaginary
for $\varepsilon>\varepsilon_{\rm cr}^{(1)}$, thus signaling the onset of the
spontaneous symmetry breaking
(SSB) instability of the dark soliton (which should be expected to
displace the
soliton from the trap center).

The dependence of $\omega_\alpha$ on the amplitude $\varepsilon$ of
the imaginary potential $W$, as found by the BdG analysis, is
illustrated in the middle and bottom panels of Fig.~\ref{fig2}.
The  mode, corresponding to the oscillatory motion of the dark soliton, is the first non-zero mode
in the real part of the spectrum, indicated with a dashed (red) line ---cf.~middle panel of Fig.~\ref{fig2}.
We observe that it initially moves towards the spectral plane origin, and past the critical point,
$\varepsilon_{\rm cr}^{(1)}$ (cf.~vertical line), $\omega_\alpha$ becomes imaginary, and thus the soliton
becomes unstable. The corresponding pair of imaginary eigenfrequencies is shown in the bottom panel of
Fig.~\ref{fig2}. The solid (black) line in both panels, shows the analytical result of Eq.~(\ref{solev}).
An excellent agreement between the analytical prediction and the BdG numerical result is observed even beyond
the SSB bifurcation point, while for larger values of the parameter $\varepsilon$, perturbation
theory fails and, as expected, the agreement becomes worse.

Importantly, for large values of the parameter $\varepsilon$, the imaginary eigenfrequencies
start moving towards the spectral plane origin and collide with it at a second critical point, $\varepsilon_{\rm cr}^{(2)}$,
after which the branch is terminated.
\begin{figure}[tbp]
\includegraphics[width=8.5cm]{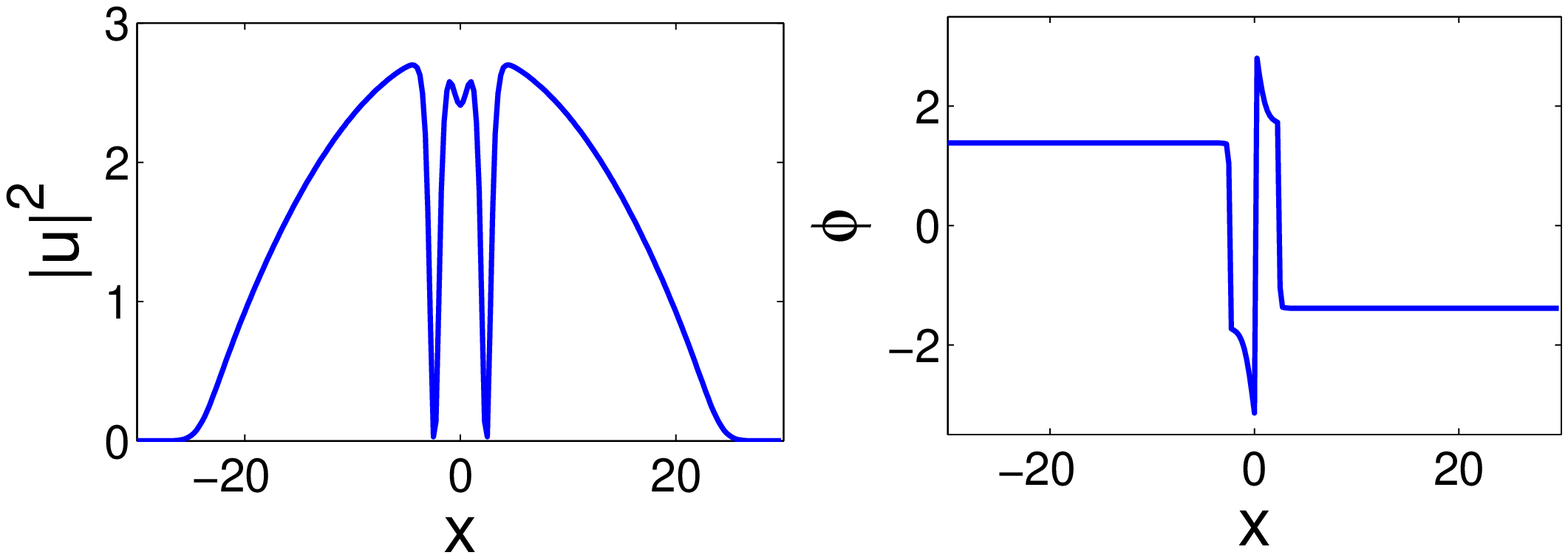}
\includegraphics[width=8.5cm]{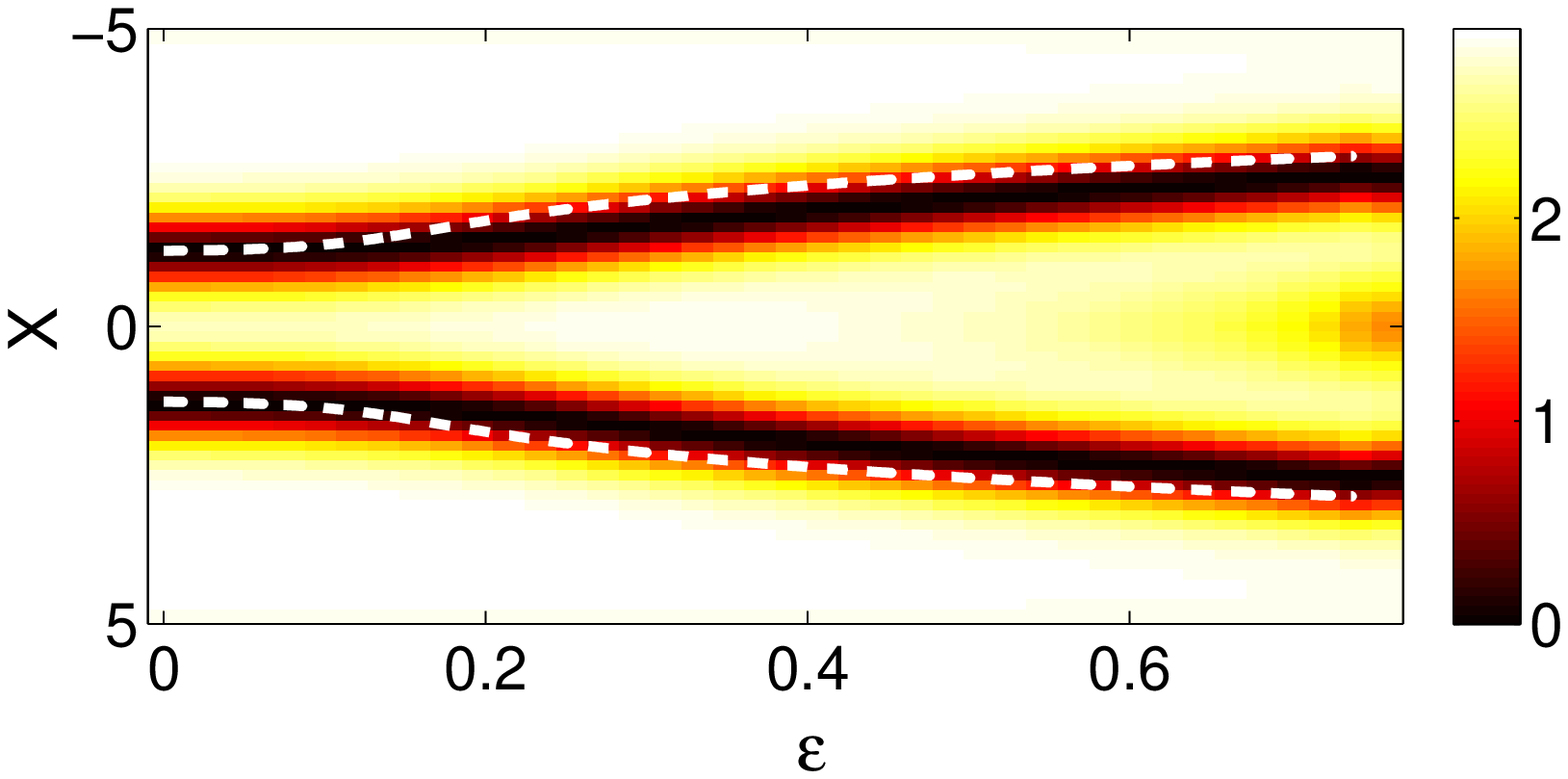}
\includegraphics[width=8.5cm]{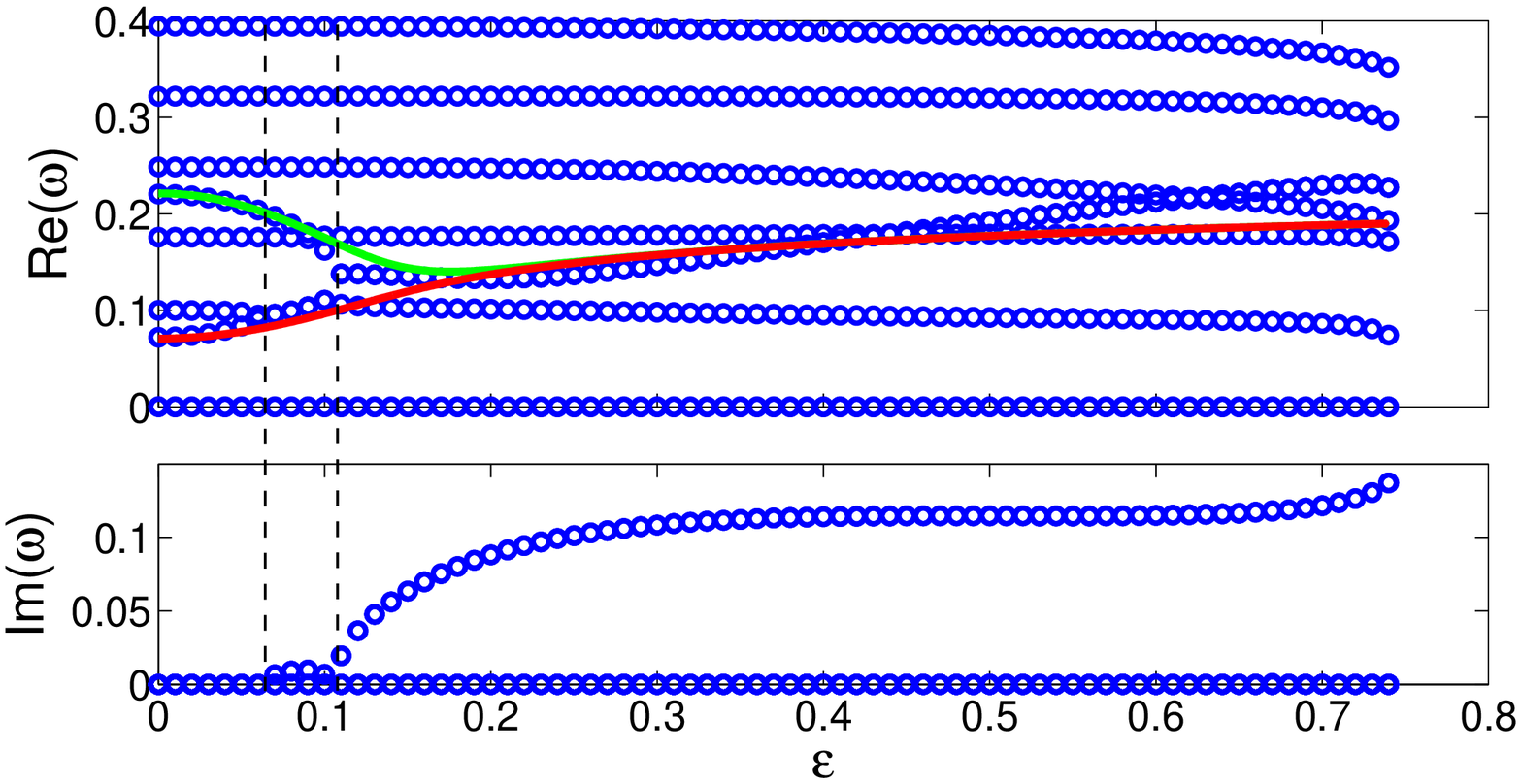}
\caption{
(Color online)  Top panels: the density (left) and phase (right) profiles of a stationary two-dark-soliton state.
Middle panel: the dependence of the equilibrium configuration of the two-soliton solution as a function of $\varepsilon$;
dashed (white) lines show the theoretical predictions for the position of the centers of the waves, cf.~fixed points
of Eqs.~(\ref{sys1})-(\ref{system1}).
Bottom panels: the linear spectrum of the two soliton branch, namely the real (top) and imaginary (bottom) parts of the eigenfrequencies as functions of $\varepsilon$. Solid red and green lines, correspond to the analytical result of Eq.~(\ref{eigs2}). The value of these modes at $\varepsilon=0$, corresponds to the first and second anomalous modes of the Hamiltonian case.
}
\label{fig23}
\end{figure}
To better understand how the branch ceases to exist, we first observe (bottom panel of Fig.~\ref{fig1}) that
the density profile of the soliton becomes increasingly shallower (i.e., more ``grey'')
as $\varepsilon$ grows and the second critical point is approached.
This is due to the development of an increasingly strong {\it even} imaginary part of the solution.
Furthermore, the
stable background solution $u_b(x)$
[cf.~Eqs.~(\ref{tf23}) and top panel of Fig.~\ref{fig1}]
develops an {\it odd} imaginary part resembling a (progressively darker) grey soliton.
Finally, at $\varepsilon=\varepsilon_{\rm cr}^{(2)}$, the profiles of these modes become
identical and disappear in a blue-sky bifurcation through their collision.
In this saddle-center bifurcation, the dark soliton plays the role
of the saddle, while the ground state of the system is the center.
This observation leads to the conclusion that there is a phase transition where the system loses its
$\mathcal{PT}$-symmetry, due to the nonlinear problem eigenvalues ceasing to
be real, in direct analogy with the linear
$\mathcal{PT}$ phase transition of Ref.~\cite{bend1}.
This will be discussed in more detail below.

However, we should highlight here that one more nontrivial question
persists in the context of the single dark soliton. Past the
critical point $\varepsilon_{\rm cr}^{(1)}$, it is expected that the instability
of the soliton at the center will provide us with ``daughter states''
which are spontaneously symmetry broken dark solitons centered at
a positive or at a negative value of $x_0$. For real chemical potentials/
propagation constants, such states can, nevertheless, not be identified.
This appears to be inconsistent with the pitchfork nature of the bifurcation
and constitutes a point to which we will return after we visit the
dark soliton evolution to appreciate the dynamical implications of this
instability.

\section{Multiple dark soliton states and nonlinear $\mathcal{PT}$ phase transitions}

In this section, we will study higher excited states, in the form of multiple dark solitons. Such states are known to exist in the presence of $V(x)$, and have been studied extensively in the context of BECs \cite{theocharis,review,dcds}. Here, we will consider them in the presence of the imaginary part of the potential $W(x)$. Particularly, we will focus
on the two- and three-dark-soliton states; qualitatively similar results can also be obtained for higher excited states.

\subsection{Two-dark-soliton state}

The density and phase of a two-dark-soliton-state are shown in the top panel of Fig.~\ref{fig23}. Similarly
to the ground state branch (see Fig.~\ref{fig1}), we observe that a density dip is formed at the center,
which becomes deeper as $\epsilon$ is increased. Additionally, we observe that the two solitons are located
away from the center, at some finite equilibrium distance $\pm X$. This equilibrium corresponds to the situation
where the repulsive force between the solitons and the
``expulsive'' (with respect to the trap center) effect, for
sufficiently large $\varepsilon$, of the
imaginary part of the potential $W(x)$ analyzed in the last
section is counter balanced by the confining nature of the real part
of the potential $V(x)$ \cite{theocharis,review}.

Below we will study analytically how the presence of the imaginary potential $W(x)$ modifies the equilibrium distance $X$ characterizing the stationary two-soliton-state, and also study small-amplitude oscillations of the solitons around their  fixed points. To obtain an approximate equation of motion [analogous to Eq.~(\ref{solev})] for the two solitons centers,
we assume that each of the two solitons feels the effective external potential of Eq.~(\ref{Veff}) and, at the same time, the solitons interact with each other. We assume that the pertinent interaction potential, $V_{\rm int}$,
for two (almost black) solitons has the form obtained from the exact two-soliton solution of the underlying
unperturbed NLS system (see details in Ref.~\cite{theocharis}), namely:
\begin{equation}
V_{\rm int}=\frac{\mu}{2\sinh^2[2\sqrt{\mu}(x_2-x_1)]}.
\label{dx01}
\end{equation}
The equation of motion for the two soliton centers (denoted hereafter by $x_1$ and $x_2$)
can now be obtained from the Lagrangian:
\begin{eqnarray}
L&=&\frac{1}{2}\dot{x_1}^2+\frac{1}{2}\dot{x_2}^2+V_{{\rm eff}}(x_1)+V_{{\rm eff}}(x_2) \nonumber \\
&-&\frac{\mu}{\sinh^2[\sqrt{\mu}(x_2-x_1)]},
\label{Lag2ds}
\end{eqnarray}
where the external effective potential $V_{{\rm eff}}$ is given in Eq.~(\ref{Veff}).
Using the Euler-Lagrange equations, and assuming that the two solitons are far away from
each other, i.e., $|x_2-x_1| \gg 0$, we approximate their mutual interaction term
in Eq.~(\ref{Lag2ds}) by its exponential asymptote and obtain the following equations of motion:
\begin{eqnarray}
\frac{d^2 x_1}{dt^2} & = &-8\mu^{3/2}e^{-2 \sqrt{\mu} (x_2-x_1)}-\tilde{\omega}^2x_1\nonumber \\
& + &\varepsilon^2c_1\sech^4(c_2x_1)\tanh(c_2x_1),
\label{sys1} \\
\frac{d^2 x_2}{dt^2} & = & 8\mu^{3/2}e^{-2\sqrt{\mu}(x_2-x_1)}-\tilde{\omega}^2x_2\nonumber \\
&+&\varepsilon^2c_1\sech^4(c_2x_2)\tanh(c_2x_2).
\label{system1}
\end{eqnarray}
where $\tilde{\omega}=\Omega/\sqrt{2}$ is the oscillation frequency of the single dark soliton in the absence of
$W$ \cite{review}. The fixed points $X_2=-X_1=X$ of the above system
can be found numerically, by setting the left-hand side equal to zero, and employing a fixed-point algorithm.

Next, we can study the stability of these equilibrium positions $X_1$ and $X_2$, by considering
small deviations, $\eta_1$ and $\eta_2$, around them.
By Taylor expanding the nonlinear terms, and keeping only linear terms in $\eta_1$ and $\eta_2$,
we derive the following set of linearized equations of motion:
\begin{eqnarray}
\!\!\!\!\!\!\!\!
\frac{d^2 \eta_1}{dt^2} & = & 16\mu^{2}e^{-4\sqrt{\mu}X}(\eta_2-\eta_1)-\tilde{\omega}^2\eta_1 \nonumber \\
&+& \varepsilon^2c_1c_2\left[5\sech^6(c_2X)-4\sech^4(c_2X) \right]\eta_1,
\label{system1linear} \\
\!\!\!\!\!\!\!\!
\frac{d^2 \eta_2}{dt^2} & = &-16\mu^{2}e^{-4\sqrt{\mu}X}(\eta_2-\eta_1)-\tilde{\omega}^2\eta_2 \nonumber \\
&+& \varepsilon^2 c_1c_2\left[5\sech^6(c_2X)-4\sech^4(c_2X) \right]\eta_2.
\label{system2linear}
\end{eqnarray}
Let us now consider the normal modes of the linearized system and seek solutions of the form $\eta_i=\eta^{0}_{i}e^{i\omega t}$, where $\omega$ is the common oscillation frequency of both dark solitons. Then, substituting this ansatz into
Eqs.~(\ref{system1linear})-(\ref{system2linear}), we rewrite the equations of motion as a matrix eigenvalue equation, namely:
\begin{equation}
-\omega^2 \mathbf{\eta}
= \left(
\begin{array}{cc}
-\tilde{\omega}^2
+f(X)
&~~ 16\mu^2e^{-4\sqrt{\mu}X}  \\[2.0ex]
16\mu^2e^{-4\sqrt{\mu}X}
&~~ -\tilde{\omega}^2+f(X)
 \\
\end{array} \right)\mathbf{\eta}.
\label{eig}
\end{equation}
where $\mathbf{\eta} =(\eta_1,~\eta_2)^T$ and the function $f(X)$ is given by:
\begin{eqnarray}
f(X)\!\!&=&\!-16\mu^2e^{-4\sqrt{\mu}X}\nonumber \\
&&\!	+\varepsilon^2 c_1c_2\left[5\sech^6(c_2X)-4\sech^4(c_2X) \right].
\end{eqnarray}
To this end, it is possible to obtain from Eq.~(\ref{eig}) the
two characteristic frequencies
\begin{eqnarray}
\omega_{1,2}&=&\sqrt{\tilde{\omega}^2+f(X)\pm 16\mu^2e^{-4\sqrt{\mu}X}},
\label{eigs2}
\end{eqnarray}
where $\omega_{1}$ ($\omega_{2}$) correspond to the in-phase (out-of-phase)
oscillations of the two solitons \cite{theocharis,review}.

The validity of the above analytical result can now be compared with the
corresponding numerical result obtained by the numerical existence theory and the BdG analysis. The latter
provides the equilibrium configuration, as well as
the real/imaginary parts of the eigenfrequencies as functions of the imaginary potential
strength $\varepsilon$, as shown in the middle and bottom panels of Fig.~\ref{fig23}, respectively.
In the middle panel, we can observe a very good agreement between the positions of the theoretically
predicted solitary wave centers [cf.~the fixed points of Eqs.~(\ref{sys1})-(\ref{system1})] and their
numerically exact counterparts.

For the linearization analysis of the bottom panels, we observe the
following.
In the Hamiltonian case, there exist two anomalous modes (and more generally, $n$ anomalous modes
for a $n$-soliton state ---see, e.g., Ref.~\cite{review}). The first, which is the lowest nonzero mode in the spectrum, corresponds to the in-phase oscillation frequency of the two dark solitons, while the second being the fourth finite eigenfrequency for our parameters, corresponds to out-of-phase oscillations.
The two-soliton branch is initially stable up to a critical point (indicated by the left vertical line), where the
first finite mode collides with the dipole (alias Kohn) mode~\cite{review}; this mode
is located, for $\varepsilon\!=\!0$, at ${\rm Re}(\omega)\!=\!\Omega\! =\!0.1$. This collision results in the emergence of an instability, which is identified by
the appearance of an imaginary part of the eigenfrequency (see the bottom panel of Fig.~\ref{fig23}). As
$\varepsilon$ is increased, this mode is eventually {\it detached} from the dipole mode and this instability band
ceases to exist for a very narrow parametric interval. However, the quadrupole mode (located, for $\varepsilon=0$,
at Re$(\omega) \approx \sqrt{3} \Omega=0.17$), which has been scattered at about $\varepsilon=0.08$, by the branch
just above, now collides with the detached mode (right vertical  line).
Upon the latter collision,
 a new (and persistent hereafter) second instability emerges (see the bottom panel of Fig.~\ref{fig23}).
This way, another quartet of complex eigenfrequencies is created, indicating
that the two-soliton state becomes unstable with a growth rate that
continues to increase as the gain/loss parameter $\varepsilon$ increases.

The analytical result for the two eigenfrequencies $\omega_{1,2}$ [cf.~Eq.~(\ref{eigs2}) and solid red and green  lines in Fig.~\ref{fig23}]
essentially coincides
with the two branches initialized at the two anomalous modes of the
Hamiltonian case, for sufficiently small $\varepsilon$;
however, for larger $\varepsilon$, the analytical result is somewhat less
accurate. The observed discrepancy is, at least in part,
due to the fact that our approximate result is based on a particular ansatz
of well-separated (thus weakly-interacting), almost black solitons. However,
observing the density profile of the two soliton state (top panel of
Fig.~\ref{fig23}), one can see that the small density dip in the center
(which is not included in our ansatz)
is certainly affecting the interaction between the two solitons;
importantly, this dip becomes larger as $\varepsilon$ is increased.
Thus, for relatively large values of $\varepsilon$, our perturbative
approach may be expected to be of lesser value.

Additionally we note that, as in the case of the single-soliton-state,
for values $\epsilon>0.72$ no stationary
two-soliton-state could be identified. This is again a by-product  of the
nonlinear analogue of the $\mathcal{PT}$-phase transition. More specifically,
the two-soliton branch collides with the three-soliton one and
they pairwise annihilate, as is occurring to higher (linear) eigenvalue
pairs in the linear $\mathcal{PT}$-phase transition of~\cite{bend1}.
We illustrate the generality of this effect (and the relevant cascade
of nonlinear eigenvalue collisions) in more detail below.

\subsection{Three-dark-soliton state}

\begin{figure}[tbp]
\includegraphics[width=8.5cm]{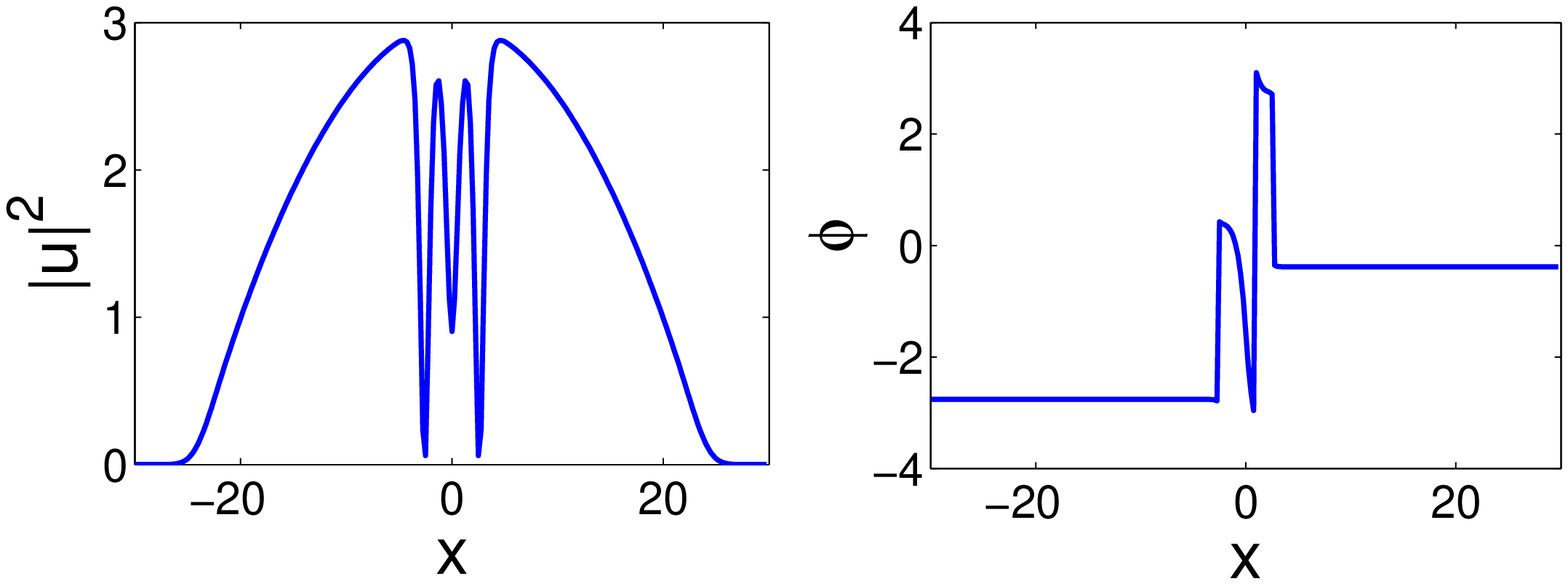}
\includegraphics[width=8.5cm]{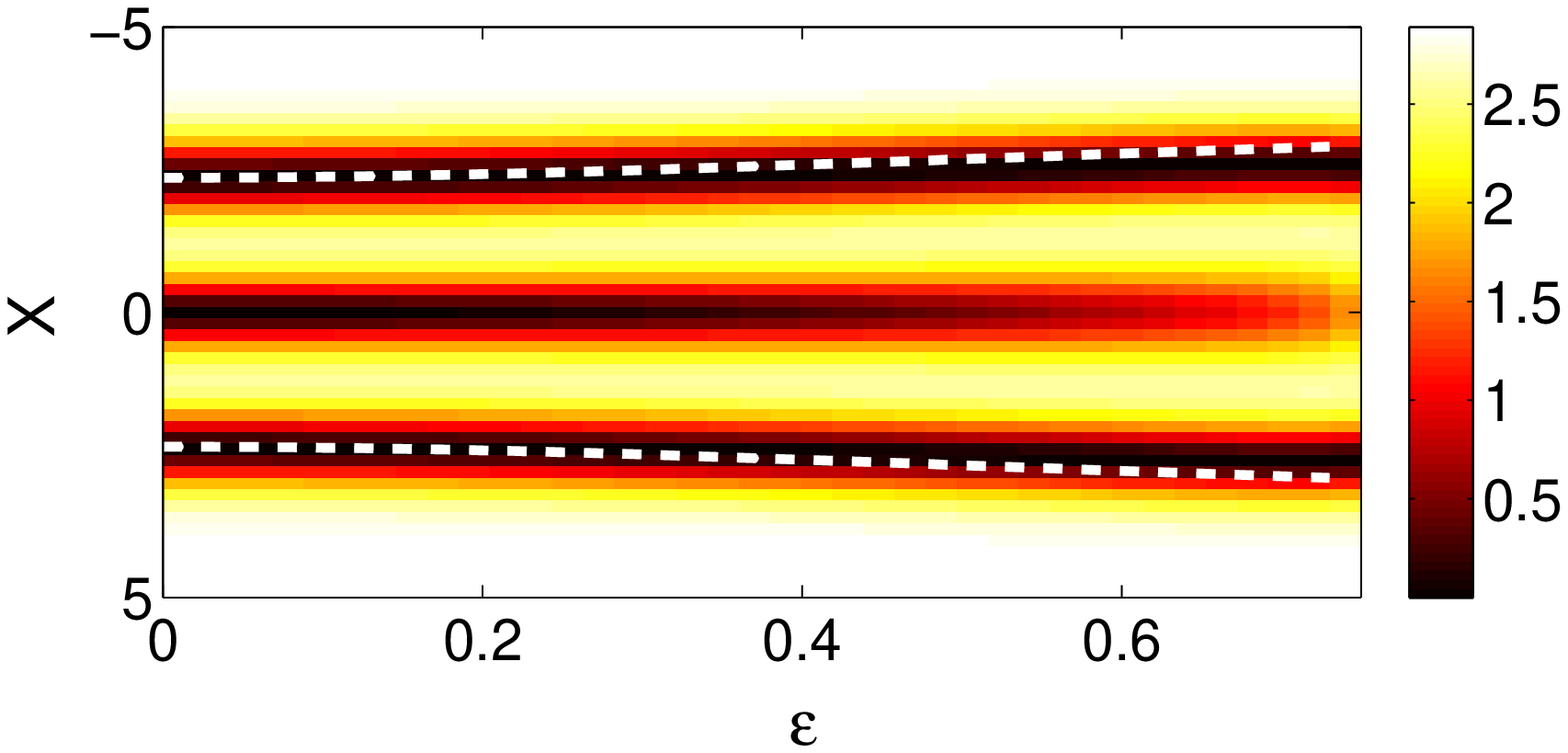}
\includegraphics[width=8.5cm]{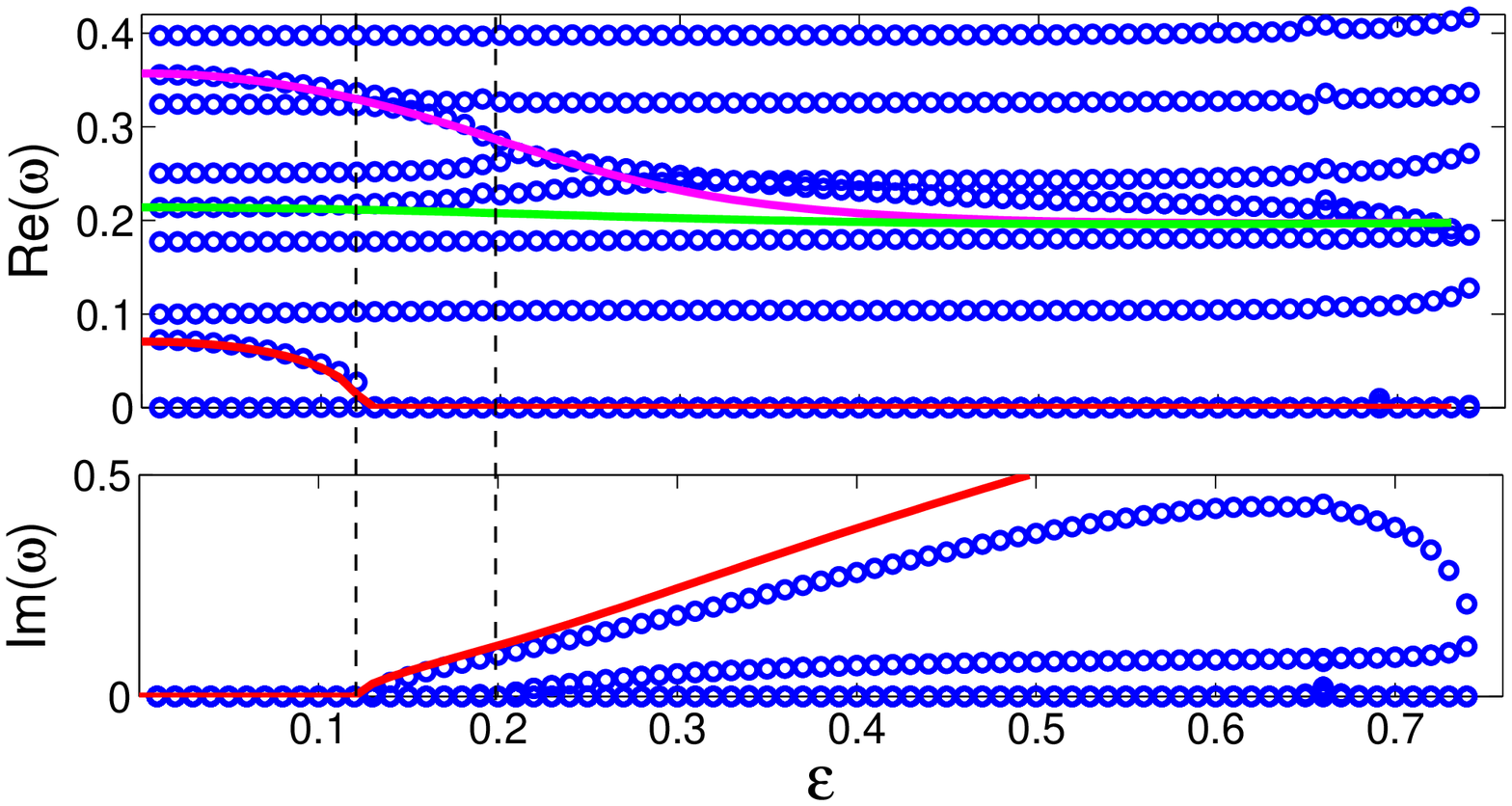}
\caption{(Color online) Same as Fig.~\ref{fig23}, but for the three soliton state.
In the bottom panels, solid red, green and magenta lines, correspond to the analytical result of Eqs.~(\ref{eigs31})-(\ref{eigs32}). The value of these modes at $\varepsilon=0$ correspond to the
first, second and third anomalous modes of the Hamiltonian case.}
\label{fig24}
\end{figure}

We now proceed with the investigation of the three-soliton branch; examples of the density and phase of such a state are shown in the top panel of Fig.~\ref{fig24}. This state shows similar behavior with the single soliton state, in the sense that the soliton located at the center becomes shallower as $\varepsilon$ is increased.

The statics and dynamics around equilibria of the three-dark-soliton state can be analyzed by the methodology adopted in the case of the two-dark-soliton state. Particularly, we will determine the equilibrium positions of the three solitons
and study their small-amplitude oscillations around their fixed points. First we note that, from symmetry arguments, we expect that the fixed points are $X_2=0$ (for the central soliton), and $X_3=-X_1=X$ (for each of the two outer solitons). For this configuration, we can again calculate the fixed points for the outer solitons numerically
and compare them to the particle theory analytical predictions, as illustrated in the middle panel
of Fig.~\ref{fig24}, obtaining good agreement for small values of $\varepsilon$.

Subsequently, we derive the linearized equations around the fixed points, and
finally obtain the three relevant eigenfrequencies of the system.
This way, we find that the frequencies of the normal modes of the system are:
\begin{eqnarray}
\omega_1&=&\sqrt{f_1(X)}
\label{eigs31}
\\
\omega_{2,3}&=&\Big\{\frac{1}{2}f_1(X)\pm f_2(X)\Big[\big(f_1(X)-f_2(X)\big)^2 \nonumber \\
 && +8f_3(X)^2 \Big]^{1/2}\Big\}^{1/2},
\label{eigs32}
\end{eqnarray}
where functions $f_{i}(X)$ ($i=1,2,3$) are given by:
\begin{eqnarray}
f_1(X)&=&\tilde{\omega}^2-16\mu^2e^{-2\sqrt{\mu}X}\nonumber
\\
&+&\varepsilon^2 c_1c_2\left[5\sech^6(c_2X)-4\sech^4(c_2X) \right],
\\
f_2(X)&=&\tilde{\omega}^2-32\mu^2e^{-2\sqrt{\mu}X}+\varepsilon^2 c_1c_2,
\\
f_3(X)&=&16\mu^2 e^{-2\sqrt{\mu}X}.
\label{fi}
\end{eqnarray}
The above characteristic frequencies will again be compared to the eigenfrequencies obtained numerically
by means of the BdG analysis. For this purpose, in the bottom panels of Fig.~\ref{fig24}, we show
the BdG linear spectrum of the three-dark-soliton branch.

As observed in the figure, the three-dark-soliton state is initially stable (for sufficiently small $\varepsilon$), but it  becomes unstable after a collision of the lowest  mode [dashed (red) line] with the spectral plane
origin, giving rise to the emergence of an imaginary pair of eigenvalues.
This is directly reminiscent of the corresponding instability of
the single dark soliton state and the mechanism of the instability is
expected to persist for {\it any} configuration with an odd number of
dark solitons in this system [we have also checked that it arises in the
case of a five-soliton state].
For larger values of $\varepsilon$, another collision between the fifth and sixth mode, results in the emergence of an imaginary eigenfrequency pair (right vertical dashed line), and thus the three soliton state remains unstable.

At the same time, and as in the single-soliton case, the imaginary eigenfrequency corresponding to the first nonzero mode, eventually returns to the origin and collides with it,
at $\varepsilon=0.72$. This is the critical point of the collision
of the two-soliton and the three-soliton branch and of their pairwise
annihilation. As the relevant  critical value of $\varepsilon$ is approached,
the center grey soliton of the three-soliton state becomes  grayer and eventually  becomes identical to the central density dip of the two-soliton state with which it collides.
The above characteristic frequencies will again be compared to the eigenfrequencies obtained numerically
by means of the BdG analysis. For this purpose, in the bottom panels of Fig.~\ref{fig24}, we show
the BdG linear spectrum of the three-dark-soliton branch.

The analytical result for the three eigenfrequencies [cf.~Eqs.~(\ref{eigs31})-(\ref{eigs32}) and solid red, green and magenta lines in Fig.~\ref{fig24}] are in a good agreement with the respective modes obtained numerically via the BdG analysis, as long as $\varepsilon$ is sufficiently small. In particular, the critical value of $\varepsilon$ where the first finite mode collides with the origin (thus giving rise to the onset of the instability) is very well predicted; furthermore, the analytically found ``trajectory'' of the other two branches follow, in a fairly good accuracy, the respective numerical result, even for moderate and large values of $\varepsilon$,
with an expected progressive discrepancy as $\varepsilon$ becomes large
(when the problem is outside the realm of a perturbative treatment).



%

\subsection{Nonlinear $\mathcal{PT}$ phase transitions}
\begin{figure}[tbp]
\includegraphics[width=7.5cm]{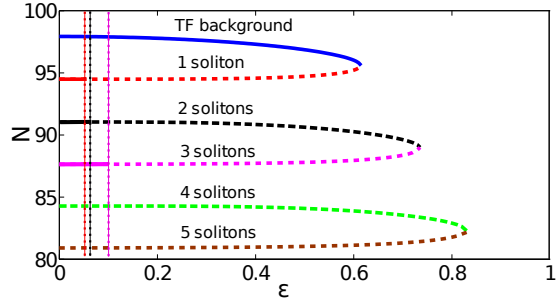}
\caption{(Color online). The power $N$ as a function of $\varepsilon$, for the six lowest states.
Solid lines correspond to linearly stable states and dashed lines to unstable states, with Im$(\omega)\neq 0$.
The vertical lines denote the critical values of $\varepsilon$ found for the single-, two- and three-dark-soliton states.
This bifurcation diagram illustrates the nonlinear $\mathcal{PT}$-phase transition points, at which stationary states disappear in pairs through
blue-sky bifurcations.}
\label{phtr}
\end{figure}

In order to further elaborate on the emergence of the nonlinear $\mathcal{PT}$-phase transition, in
Fig.~\ref{phtr} we plot the power $N$ as function of the parameter $\varepsilon$, for the lowest six (ground state and the first five excited) states of the system. The top solid (blue) branch depicts the stable ground state [i.e., the TF background $u_b$ ---cf.~Eqs.~(\ref{TF1})-(\ref{tf23})], which ultimately collides with the single-dark-soliton branch [dashed (red) line], i.e., the first excited state at $\varepsilon\approx 0.62$ (for $\mu=3$). Furthermore, the next pair of branches, corresponding to the two- and three-dark-soliton states (depicted by the dashed black and magenta lines, respectively), also collide at $\varepsilon\approx 0.72$. This picture of colliding and annihilating pairs of excited states remains the same even for higher excited states ---cf.~the lowest dashed (green and brown) curves in Fig.~\ref{phtr}, showing the annihilation of the five- and six-dark-soliton states as well.

An important general remark, regarding the structure of the bifurcations shown in Fig.~\ref{phtr}, is that higher excited states sustain these collisions,
saddle-center bifurcations and corresponding blue-sky annihilations
for progressively larger values of $\varepsilon$. This
feature motivates us to refer to this process as the {\it nonlinear analogue} of the ``traditional'' $\mathcal{PT}$ transition, in direct correspondence
with the pairwise collisions in Ref.~\cite{bend1}
(see, e.g., Fig.~1 of that reference) for the linear setting. Although
the two figures are similar, there are some interesting differences.
One such is that in the
linear picture the lowest eigenvalues collide the last, while in the 1D nonlinear case the lowest eigenstates disappear first~\footnote{Another important
difference that we do not touch upon here is that in the model
of Ref.~\cite{bend1},
the ground state does not collide with some other linear mode of the system,
contrary to what is the case herein.}. This feature has important
consequences for the supercritical evolution of the system which we will
revisit in the dark soliton dynamics section below.

\subsection{The ``free space'' case}

In our previous considerations, we have studied the case where the real part of the external potential was parabolic.
Nevertheless, and as a complementary setting,
it is also interesting to consider the ``free space'' case, where the real trapping potential
is absent, i.e., $V(x)=0$. Such a situation may occur, e.g., in the context of optics where ---instead of a graded index medium--- one may consider a medium with a constant linear refractive index, i.e., $V(x)=V_0={\rm const.}$; in this case, the pertinent term $V_0 u$ that would appear in Eq.~(\ref{PT1}) can straightforwardly be removed by a trivial gauge transformation.

An interesting feature characterizing the free space case is that stationary multiple dark soliton states can no longer exist: in this case, the repulsion between dark solitons cannot be counter balanced by any restoring force
as the $V(x)$ producing the latter is absent.
As a result, the most fundamental states pertinent to this setting are
the ground state and the single-dark-soliton state. An approximate solution for the free-space ground state $u_0$, can be found employing Eq.~(\ref{tf23}) and setting $V(x)=0$, namely:
\begin{eqnarray}
u_0&=&\mu-\frac{1}{\sqrt{\mu}} {\cal W}^2\,\exp\left(2i \int {\cal W}\, dx \right).
\label{tf24}
\end{eqnarray}
Using the above result, and following the analysis of the previous sections,
we find that the evolution of the dark soliton center is described by [cf.~Eq.~(\ref{solev})]:
\begin{equation}
\frac{d^{2}x_{0}}{dt^{2}}=\frac{6}{5} \varepsilon^2 x_0.
\label{solevh}
\end{equation}
The above equation dictates that a quasi-stationary (almost black) dark soliton is always unstable for {\it any finite} $\varepsilon$, since it experiences an effective expulsive force, which tends to displace it from the origin and
set it into motion with an increasing (magnitude of) acceleration, at least
for small values of $x_0$.
This is in contrast to the case of a parabolic $V(x)$, where the respective force exerted on the soliton and the effective expulsive force induced by the
imaginary potential $W(x)$ could balance each other.
Thus, the presence of the real parabolic trap (or, more generally, of a
confining potential) is crucial for the existence of stable
stationary soliton solutions.
\begin{figure}[tbp]
\includegraphics[width=8.0cm]{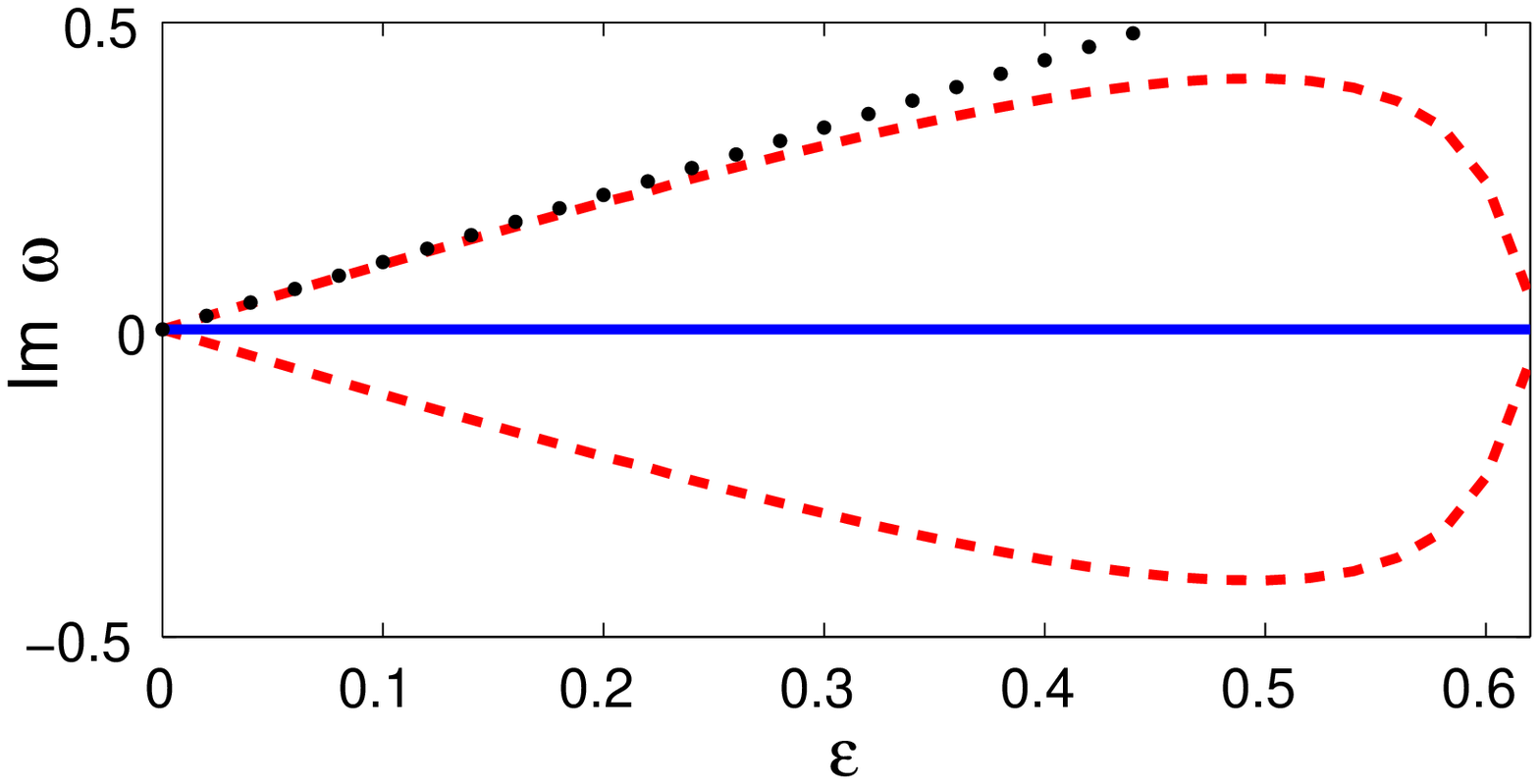}
\includegraphics[width=8.0cm]{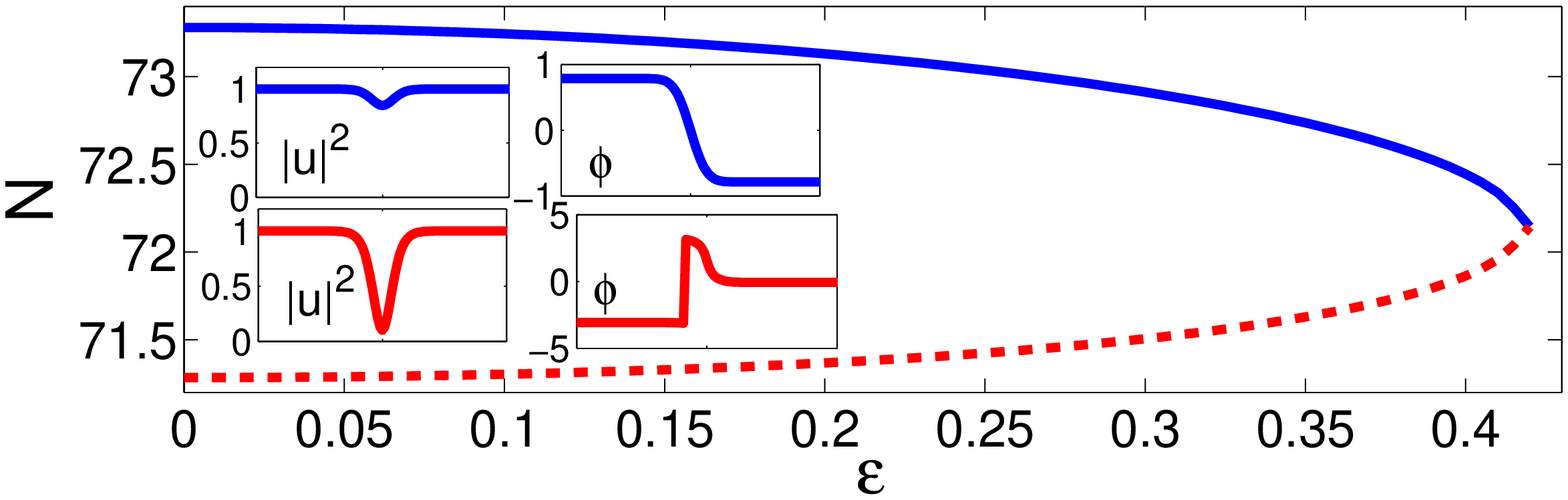}
\caption{(Color online) Top panel: pertinent eigenfrequency of
the linear spectrum for the free-space dark soliton with respect to
$\varepsilon$. Dashed (red) line depicts the imaginary eigenfrequency pair
arising due to the presence of $W(x)$ and the dotted line represents the
corresponding theoretical prediction.
Bottom panel: the $\mathcal{PT}$-phase transition diagram in the $(N,\epsilon)$ plane. Solid (blue) line shows the
background branch, while dashed (red) line the single-dark-soliton state. The upper and lower insets show characteristic
densities and phases for the ground state and the dark soliton, respectively. In this case, $\mu=1$.}
\label{phtrhom}
\end{figure}

The top panel of Fig.~\ref{phtrhom} shows
the relevant eigenfrequency for the
linear stability spectrum of the dark soliton, with respect
to the parameter $\varepsilon$. As predicted by Eq.~(\ref{solevh}), as soon
as $\varepsilon$ becomes non-vanishing, the dark soliton becomes
unstable. Its spectrum shows a pair of imaginary eigenfrequencies emerging
from the spectral plane origin in good agreement (for small/intermediate
values of $\varepsilon$) with the corresponding theoretical prediction.
Similarly to the case with the real parabolic
potential (see Fig.~\ref{fig2}), this pair of eigenfrequencies eventually
returns to ---and collides with--- the spectral plane origin, signaling the
termination of the branch and the onset of the $\mathcal{PT}$-transition.
In the bottom panel of Fig.~\ref{phtrhom}, the bifurcation diagram for the
ground state and the
single-dark-soliton  in the free-space case is plotted; we observe a behavior similar to the one found in the
case with the parabolic $V(x)$, but with the major difference that beyond the
critical point
($\varepsilon=0.4$ in this case) {\it no stationary states} exist, as the
only ones such have collided and disappeared in the saddle-center bifurcation
discussed above.
%
\begin{figure}[t]
\includegraphics[width=4.2cm]{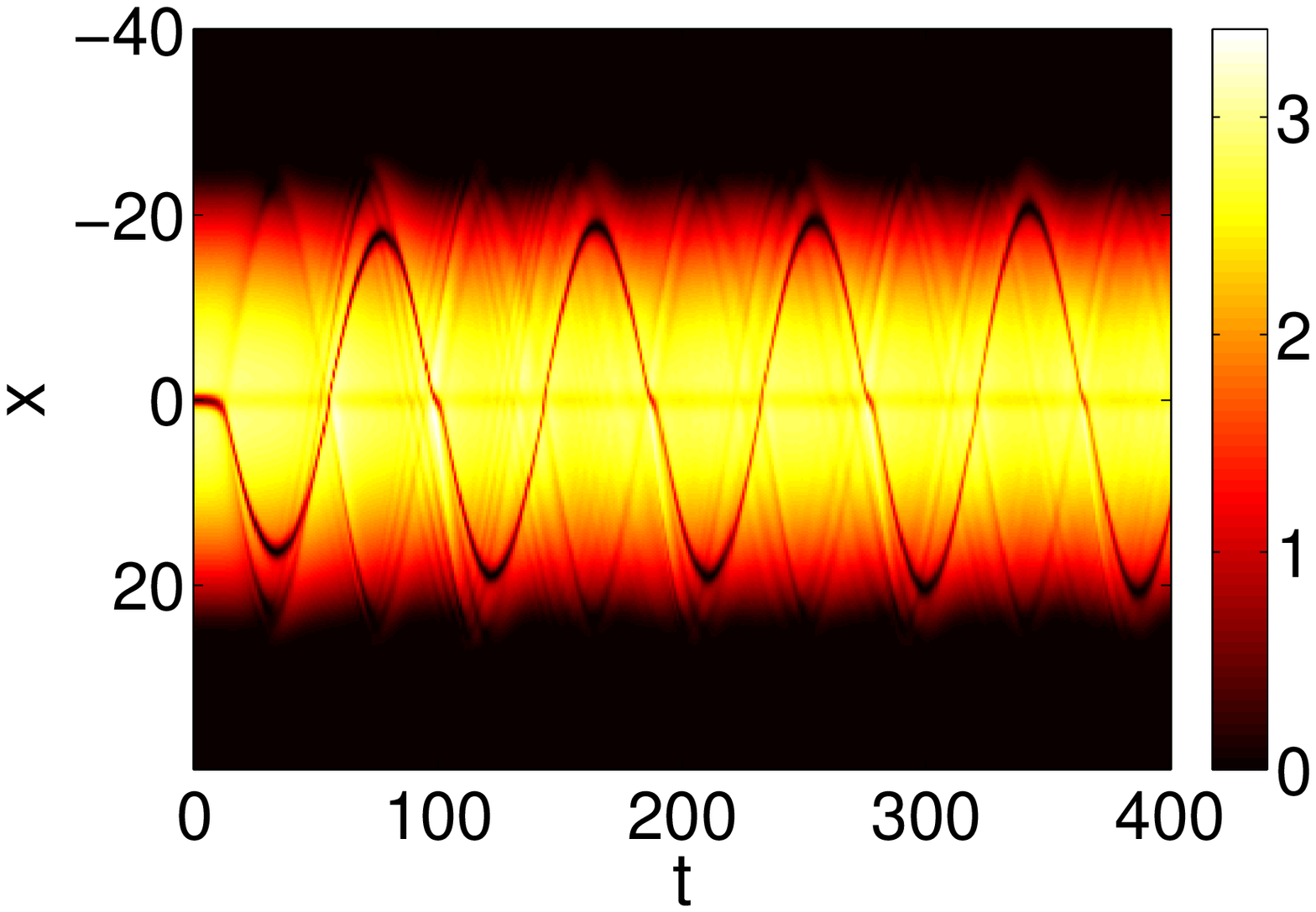}
\includegraphics[width=4.2cm]{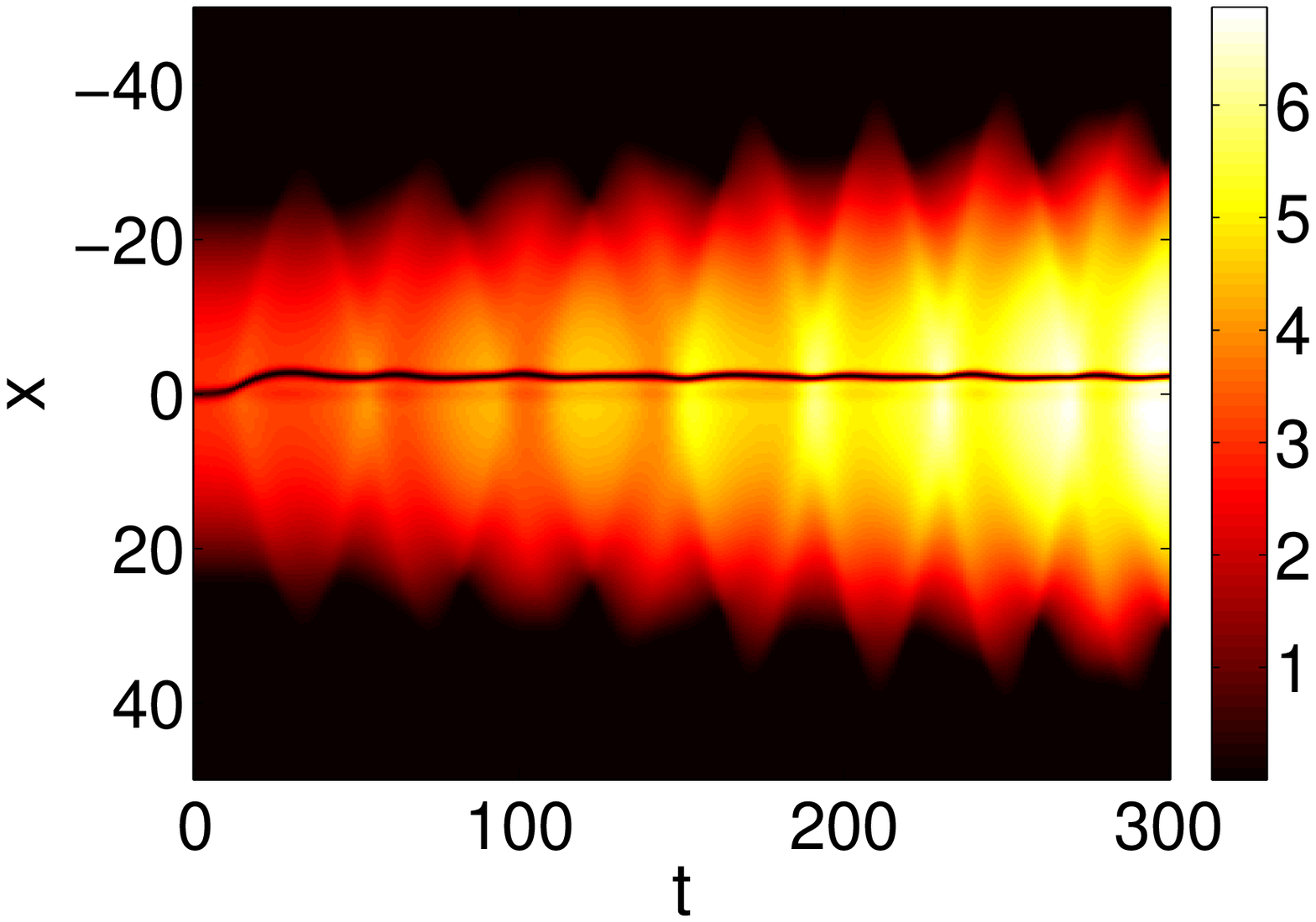}
\includegraphics[width=4.2cm]{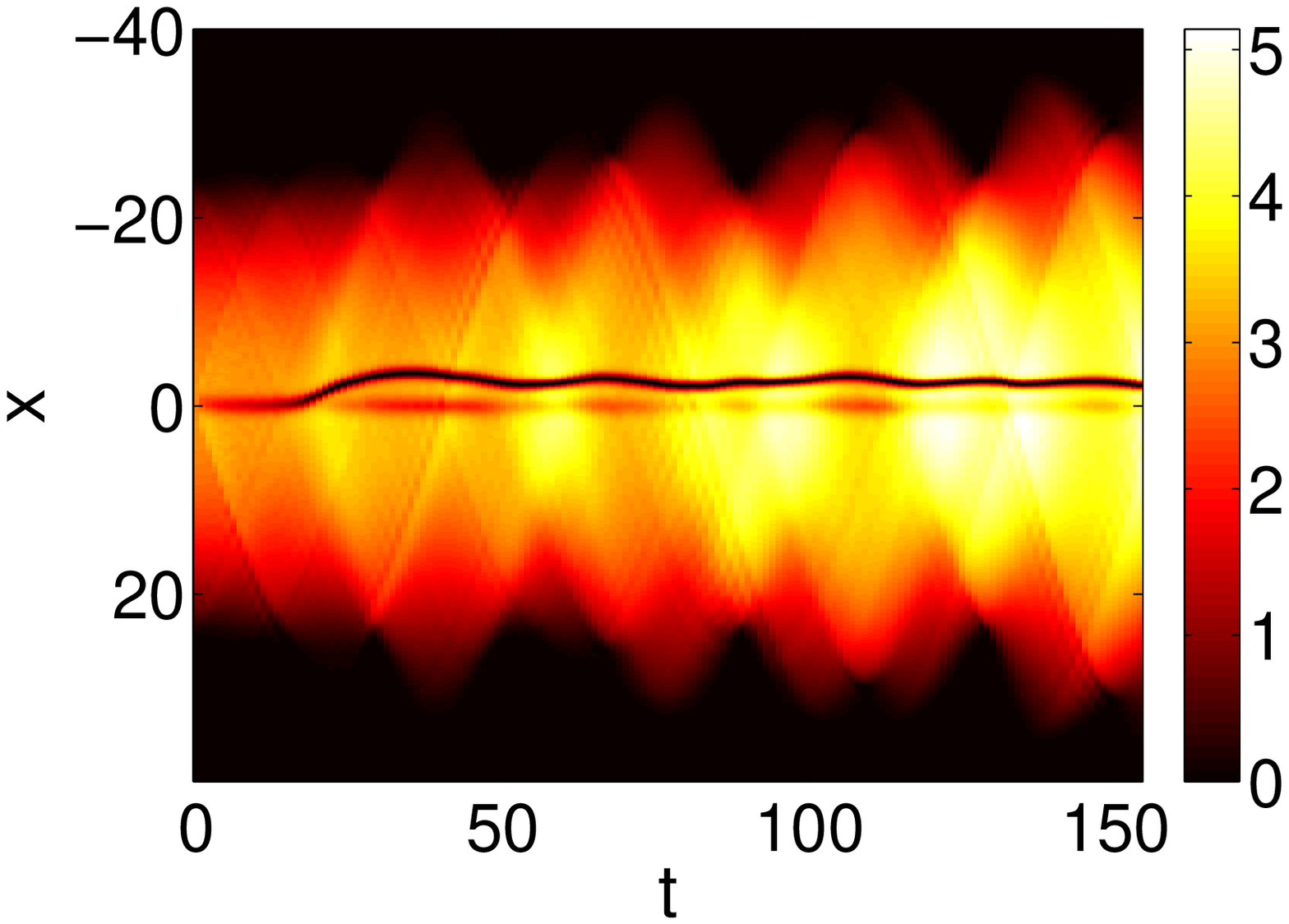}
\includegraphics[width=4.2cm]{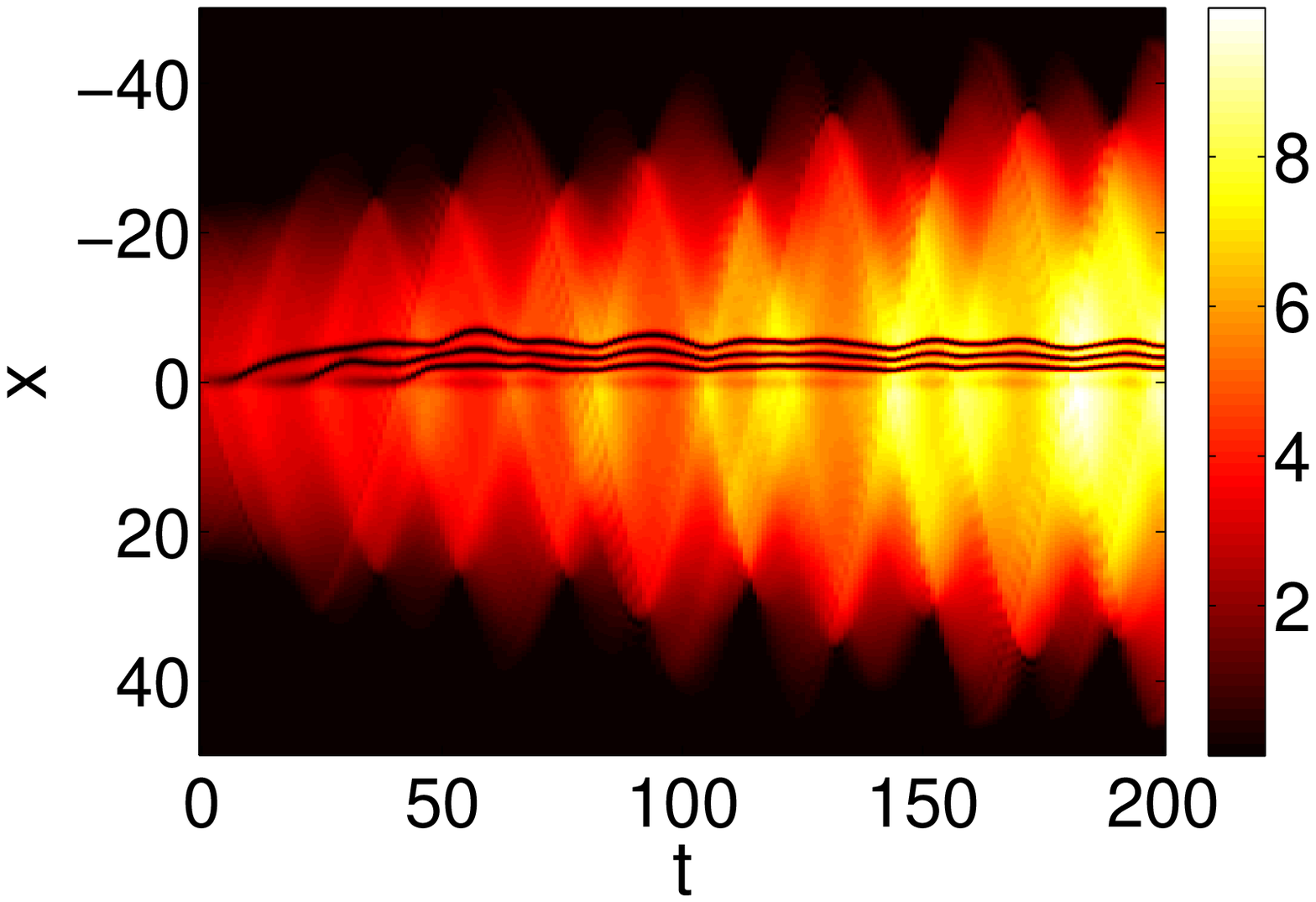}
\caption{(Color online) Bifurcation-induced dynamics. Top panels:
two potential
manifestations of the SSB destabilization scenarios for a single unstable dark
soliton past $\varepsilon=\varepsilon_{\rm cr}^{(1)}$. Bottom panels:
soliton ``sprinklers'' spontaneously leading to two or  four solitons from
the ground state (initial condition)
used for  $\varepsilon>\varepsilon_{\rm cr}^{(2)}$.
The parameters are $\mu=3$ and $\Omega=0.1$ and $\varepsilon=0.4$
(top row), $\varepsilon=0.63$ (bottom left), and $\varepsilon=0.66$
(bottom right).}
\label{fig3}
\end{figure}	
%
\section{Dark soliton dynamics}
%
We continue by considering the dynamics of the nonlinear excited states,
by numerically integrating Eq.~(\ref{PT1}) in various different regimes in $\varepsilon$.
To be more specific, our aim is to illustrate the dynamics of the dark soliton upon its destabilization
after the SSB bifurcation at $\varepsilon_{\rm cr}^{(1)}$, and also beyond the nonlinear $\mathcal{PT}$-phase transition.
Generally, to study the dynamics of unstable solitons, we prepare a stationary dark soliton solution
in the unstable regime (dashed parts of the soliton branches in Fig.~\ref{phtr}), we add
a small random perturbation, and then allow the resulting configuration to evolve in time.
We have found that the dynamics of the system strongly depends on whether the soliton
will be displaced ---as a result of the instability--- towards $x>0$ [the ``gain side'' of $W(x)$] or
$x<0$ [the ``loss side'' of $W(x)$].

First, we consider a case where a dark soliton is spontaneously ejected
to the gain side of the imaginary potential; a pertinent example is shown in
the top left panel of Fig.~\ref{fig3}. It is observed that, in this case,
the solitary wave
starts to perform oscillations of large amplitude, a behavior which is generic for
all solitons that are initially located at $x>0$.
On the other hand, if the soliton is initially kicked towards
the ``lossy'' side of $W(x)$, upon an initial displacement of
a finite $\varepsilon$-dependent size,
it stops moving and remains quiescent,
while the background on which it ``lives''
begins to grow in amplitude and width.
A relevant example is illustrated in the top right panel of Fig.~\ref{fig3}.
We once again highlight the apparent incompatibility of the above scenaria
with the expectation of a double well effective potential for the soliton
and the SSB bifurcation of the soliton at $x_0=0$. This paves the way for
the examination of the ghost states in the next section.

The above two cases are the prototypical possibilities (independently of
$\varepsilon$) for the fate of the dark soliton upon its destabilization,
when the dark soliton state still exists. We now turn to the examination of
what happens beyond $\varepsilon=\varepsilon_{\rm cr}^{(2)}$
[where the TF background
and the single soliton are no longer stationary solutions of
Eq.~(\ref{PT1})]. There, we use, e.g., an initial
condition in the form of a TF background $u(x,0)=\sqrt{\mu-V(x)}$, pertinent to the conservative
system~\cite{emergent} as a means of exploring how the system
responds when its fundamental states are no longer present.
We have found that
a dark soliton train is spontaneously formed, with an increasingly larger
number of solitons as larger values
of $\varepsilon$ are employed.
The bottom panels of Fig.~\ref{fig3} depict a couple of examples
of this phenomenon. This constitutes a form
of what can be dubbed a ``soliton sprinkler''.
This can, at least in part,
also be intuitively connected to the observation of Fig.~\ref{fig2}
that higher excited multi-soliton states persist for larger
$\varepsilon$ than lower ones.
Nevertheless, it should also be highlighted that the observations typically
suggest that the solitons are nucleated and stay in the vicinity of the
global minimum of $W(x)$, and more generally tend to become
stationary residing on the ``lossy'' side of the imaginary potential,
similarly to the dynamical state observed above and accompanied by the
same kind of background growth.


\begin{figure}[t]
\includegraphics[width=4.2cm]{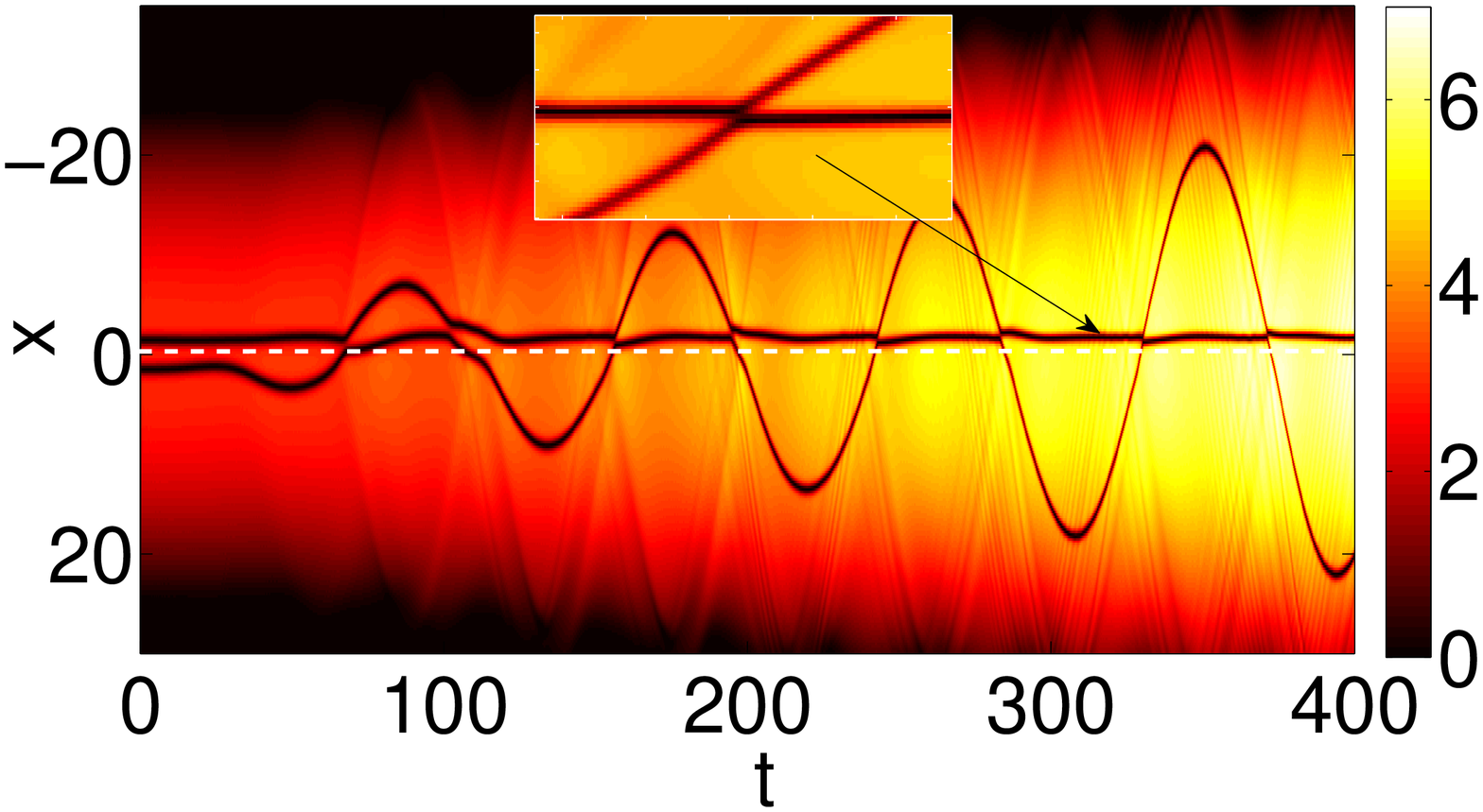}
\includegraphics[width=4.2cm]{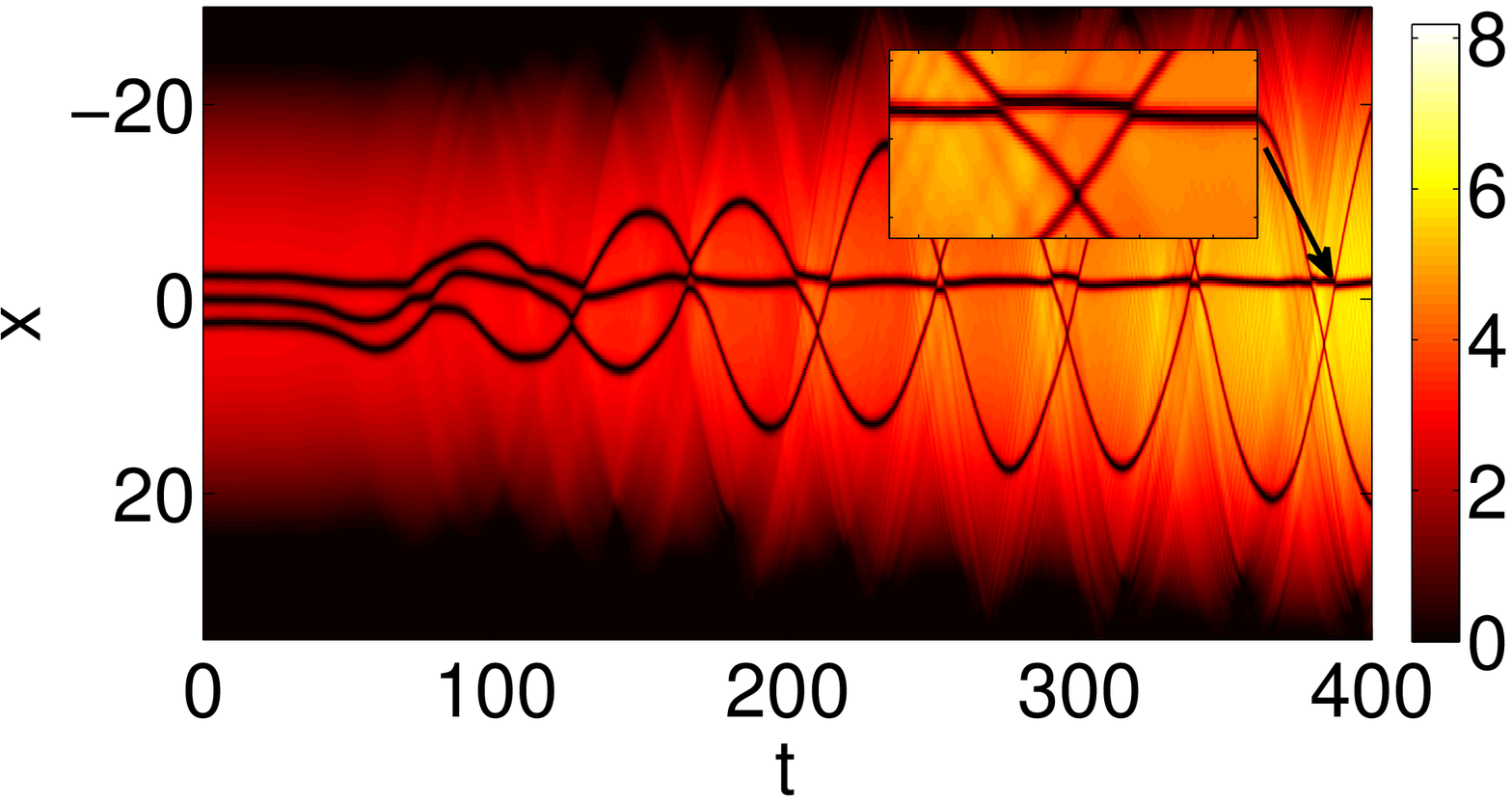}
\includegraphics[width=4.2cm]{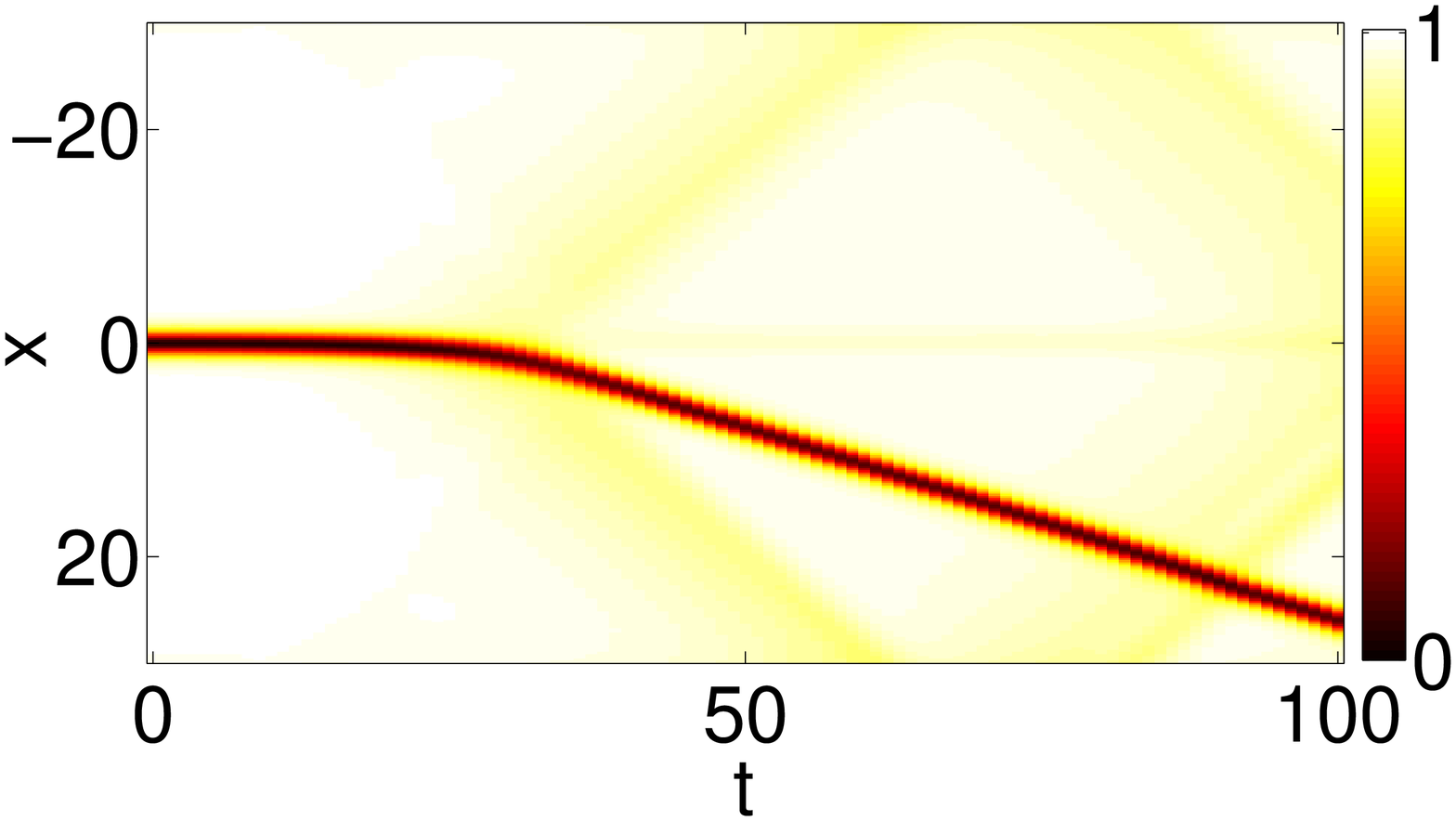}
\includegraphics[width=4.2cm]{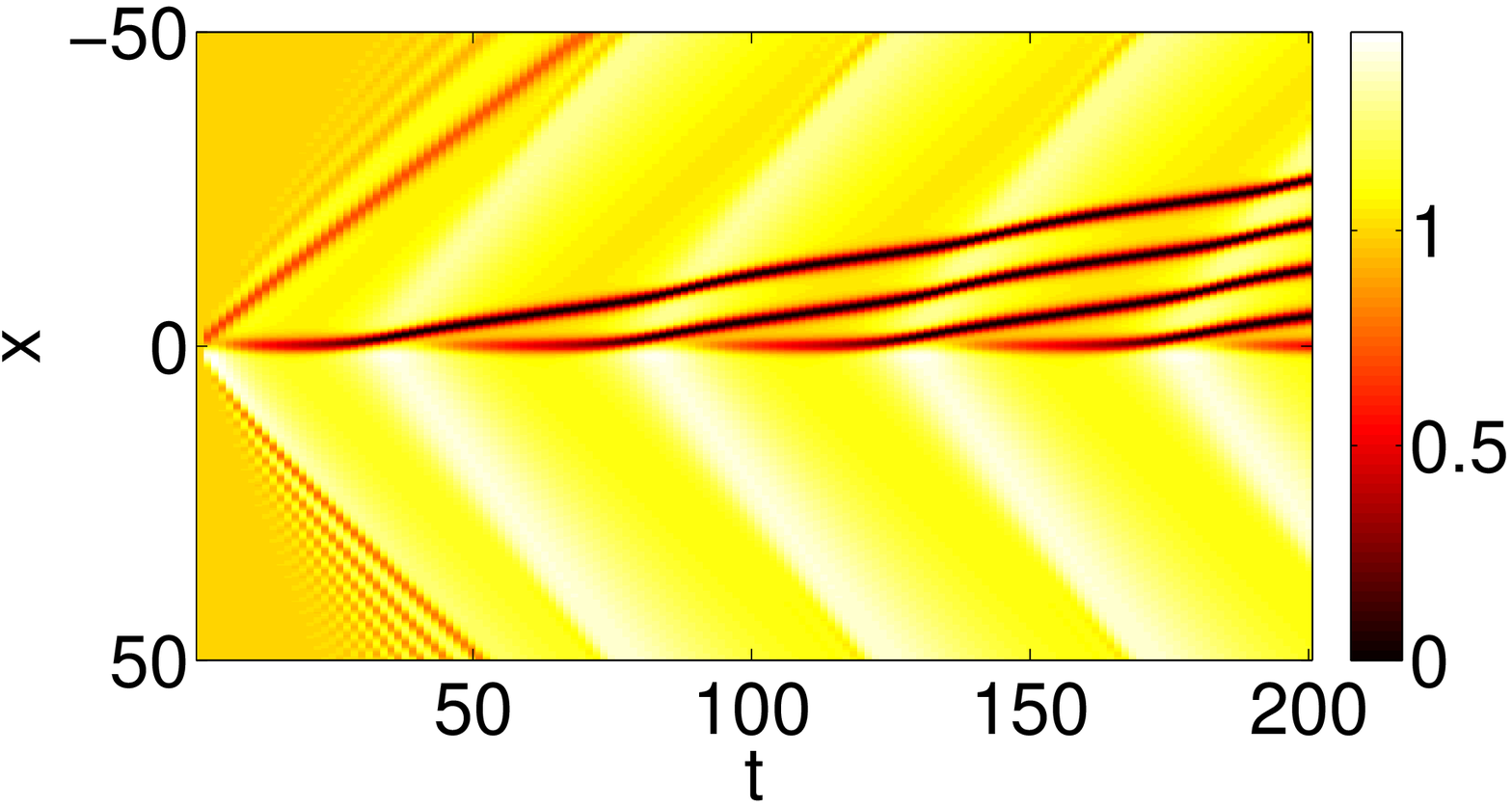}
\caption{(Color online) Bifurcation-induced dynamics. Top panels:
Manifestation of the dynamical instability of the two soliton
state (oscillatory instability of the top left panel) and
of the three soliton state (exponential and oscillatory instability
of the top right panel). Both end up with one solitary wave sitting
on the lossy side and the rest impinging on it through collisions
between turning points within a continuously growing background.
The bottom panel shows the case without $V(x)$ where the single soliton
manifests its instability by being set into motion (bottom left panel),
while a uniform initial condition past the  $\varepsilon_{\rm cr}^{(2)}$
starts emitting dark solitons in the form of a sprinkler.
The parameters are $\mu=3$ and $\Omega=0.1$ and $\varepsilon=0.4$
(top left), $\varepsilon=0.4$ (top right), while $\mu=1$ and $\varepsilon=0.1$
(bottom left), and $\varepsilon=0.45$ (bottom right).}
\label{fig33}
\end{figure}

As an additional aspect of the
investigation of  the dynamics of higher-excited (multi-soliton) states, and
how their
instabilities manifest themselves
when they are subject to small perturbations,
we consider the runs of the top panel of Fig.~\ref{fig33}.
In the left panel, the dynamics of a two-soliton state is shown.
At $t=0$ the two solitons are in equilibrium, but the soliton on the gain
side starts to execute oscillations within the trap, as was
the case before in Fig.~\ref{fig3} for the soliton displaced towards the
gain side (top left panel of the figure). Thus,  inevitably, this soliton,
upon hitting a turning point in its oscillation, returns and
collides with the other one. The result of the multiple collisions that
ensue is that one
of the solitons continues to execute oscillations of increasing amplitude
while the other one remains effectively stationary on the lossy side.
While the above sequence of oscillations and collisions is continuously
repeated, at the same time, the background grows both in amplitude and width.
It is interesting to note that the two solitons behave in a way that resembles a combination of the two different scenaria observed in the single soliton case, with the relevant dynamics resembling a ``superposition'' of the ones shown in the top two panels of Fig.~\ref{fig3}. Yet there is one non-trivial difference.
While the amplitude of oscillation in the left panel of Fig.~\ref{fig3}
remains roughly unaltered, the one of the top left panel of Fig.~\ref{fig33}
keeps increasing due to the continuously expanding background
(which, in turn, shifts the turning point of the corresponding oscillation).

A similar behavior is also found in the case of three solitons, as shown in the top right panel of Fig.~\ref{fig33}.
Two of the three solitons are now oscillating inside the harmonic potential while the third one is trapped in the lossy side of the potential.
As the solitons pass through
the origin,
they perform their respective oscillations
and undergo collisions, in a procedure which is continuously repeated; at the same time, once again, the background is growing and this, in turn, leads to
a growth of the turning points of the nearly alternating oscillations
of the two non-stationary solitons.

We now compare the above results to the ones corresponding to the free-space case, i.e., when the trapping potential $V(x)$ is absent. Recall that in this case only two stationary solutions exist (the ground state and the single soliton) and the soliton is always unstable. Since the background wave is now of constant density, when the soliton is perturbed, it will start to move away from the origin and will accelerate according to Eq.~(\ref{solevh})
up to the point where it is practically no longer under the influence of the
(exponentially localized) imaginary part of the potential.
Thereafter, it will acquire a nearly constant velocity and will keep moving
in that direction,
as shown in the bottom left panel of Fig.~\ref{fig33} (the absence of
the real part of the potential leads to a lack of a turning point in this
case). In the absence of $V(x)$, the soliton dynamics does
not depend on which direction (towards $x<0$ or $x>0$) the soliton
will be initially displaced. An initial kick in the direction of
$x<0$ will produce a similar phenomenology but for negative values of
$x$.

Finally, we consider the parameter regime
$\varepsilon > \varepsilon_{\rm cr}$, i.e., after the occurrence of the $\mathcal{PT}$-phase transition where no stationary solutions exist. We numerically integrate Eq.~(\ref{PT1}), with an initial condition of the form of a plane wave $|u(x,0)|=\sqrt{\mu}$, and the result is shown in the right bottom panel of
Fig.~\ref{fig33}. The initial plane wave begins to emit dark solitons, which are all starting to propagate along the ``lossy'' direction. Contrary to the
confined case where the trapping potential was keeping the solitons near the
origin, the solitons are continuously created and are now free to propagate along the $x$ axis.
Hence, once the ``sprinkling'' process begins, it continues indefinitely.

\section{Ghost States and their Dynamical Role}

It is now relevant to try to address some of the remaining questions,
concerning the nature of the daughter states within the pitchfork
bifurcation (of Fig.~\ref{fig2} and by extension e.g.~in Fig.~\ref{fig24})
and the unexpected asymmetric behavior of the dynamical evolution of
unstable dark solitary waves in Fig.~\ref{fig3}. A related issue
concerns the
feature that solitary waves become stationary on the lossy side
of the potential (while the corresponding background starts growing);
see Fig.~\ref{fig3} and by extension also Fig.~\ref{fig33}.

A first way to try to address this problem is to consider the possibility
(as suggested by the direct numerical simulations) of a state involving
a stationary dark solitary wave off of $x_0=0$ (and, in particular,
for $x_0<0$). Then the steady state problem would read:
\begin{eqnarray}
\mu u = {\cal L} u + i W(x) u + |u|^2 u,
\label{chpot1}
\end{eqnarray}
where ${\cal L}$ is used to denote the linear, Hermitian part of the
right hand side operator. The corresponding conjugate equation then reads:
\begin{eqnarray}
\mu^{\ast} u^{\ast} = {\cal L} u^{\ast} - i W(x) u^{\ast} + |u|^2 u^{\ast}.
\label{chpot2}
\end{eqnarray}
Now, by multiplying Eq.~(\ref{chpot1}) by $u^{\ast}$, and
Eq.~(\ref{chpot2}) by $u$, integrating and subtracting the second
equation from the first, we obtain the self-consistency
condition for the imaginary part of the chemical potential:
\begin{eqnarray}
{\rm Im}(\mu)=\frac{1}{N}\int W(x) |u|^2 dx.
\label{chpot3}
\end{eqnarray}
Interestingly, the right hand side is directly related  to the one arising
in Eq.~(\ref{dNdt}) and, thus, also controls the time dependence
of the atom number/optical power (or mathematically the $L^2$ norm).

The key realization that emerges from Eq.~(\ref{chpot3}) is that
since $|u|^2$ is an even function for solitary structures centered
at $x_0=0$ (and $W(x)$ is anti-symmetric), the only way for our states
to be centered at $x_0 \neq 0$ is if $\mu_I \equiv {\rm Im}(\mu) \neq 0$.
However, this is an atypical feature for states arising in the
context of the nonlinear Schr{\"o}dinger (NLS) equation. More specifically,
a fundamental underlying assumption behind the identification of
stationary states, critically employing the U$(1)$ invariance of
the model, is the use of the product ansatz for stationary states
of the form $u(x,t)=e^{-i \mu t} f(x)$. Now, we find ourselves
at a junction where solutions may exist with a complex $\mu$.
This, more specifically, implies that they will be indeed
exact solutions of the {\it stationary} NLS problem, yet their
incompatibility with the above ansatz {\it precludes} them from
being exact solutions of the original dynamical problem of
Eq.~(\ref{PT1}).

\begin{figure}[t]
\includegraphics[width=4cm]{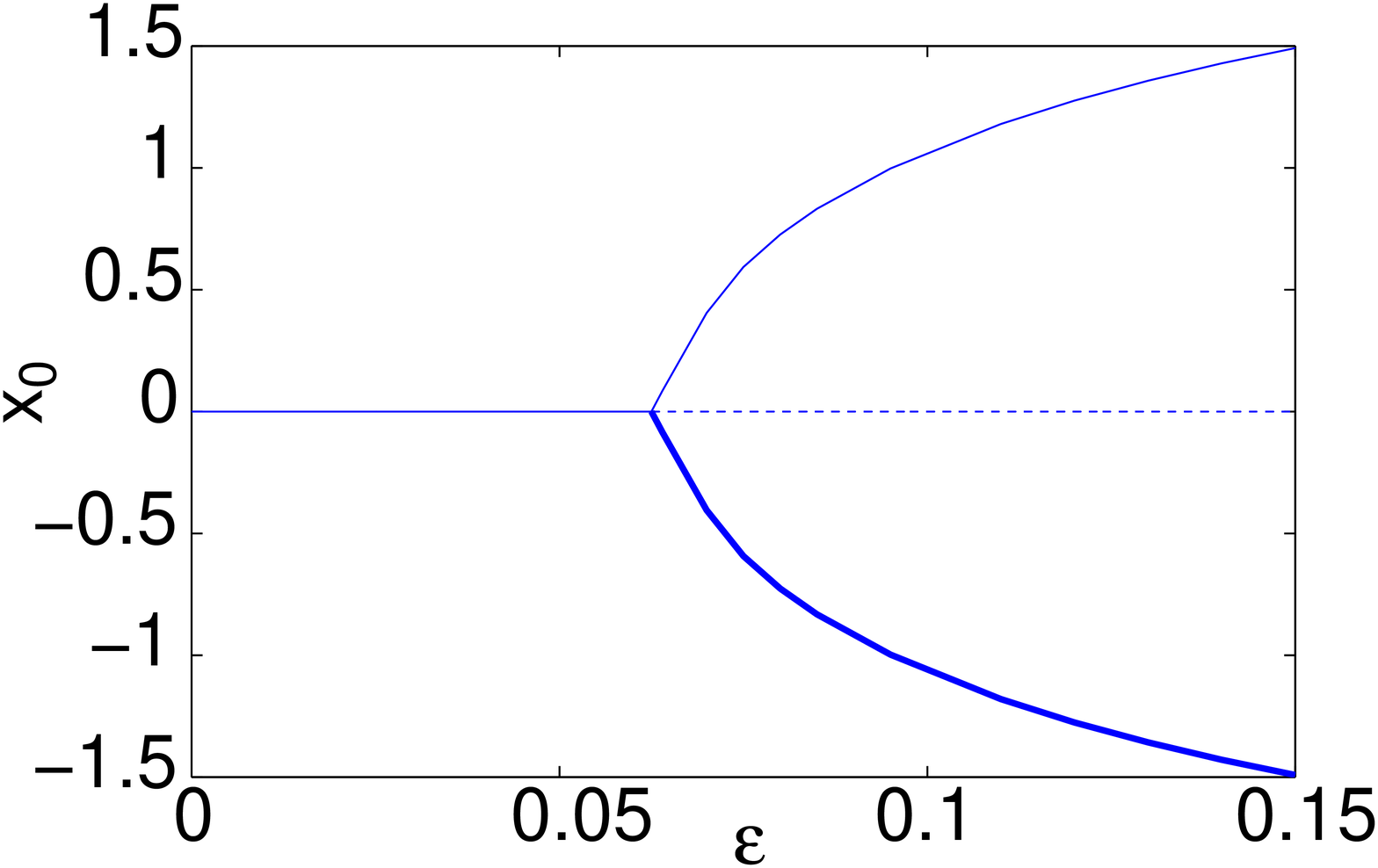}
~\includegraphics[width=4cm]{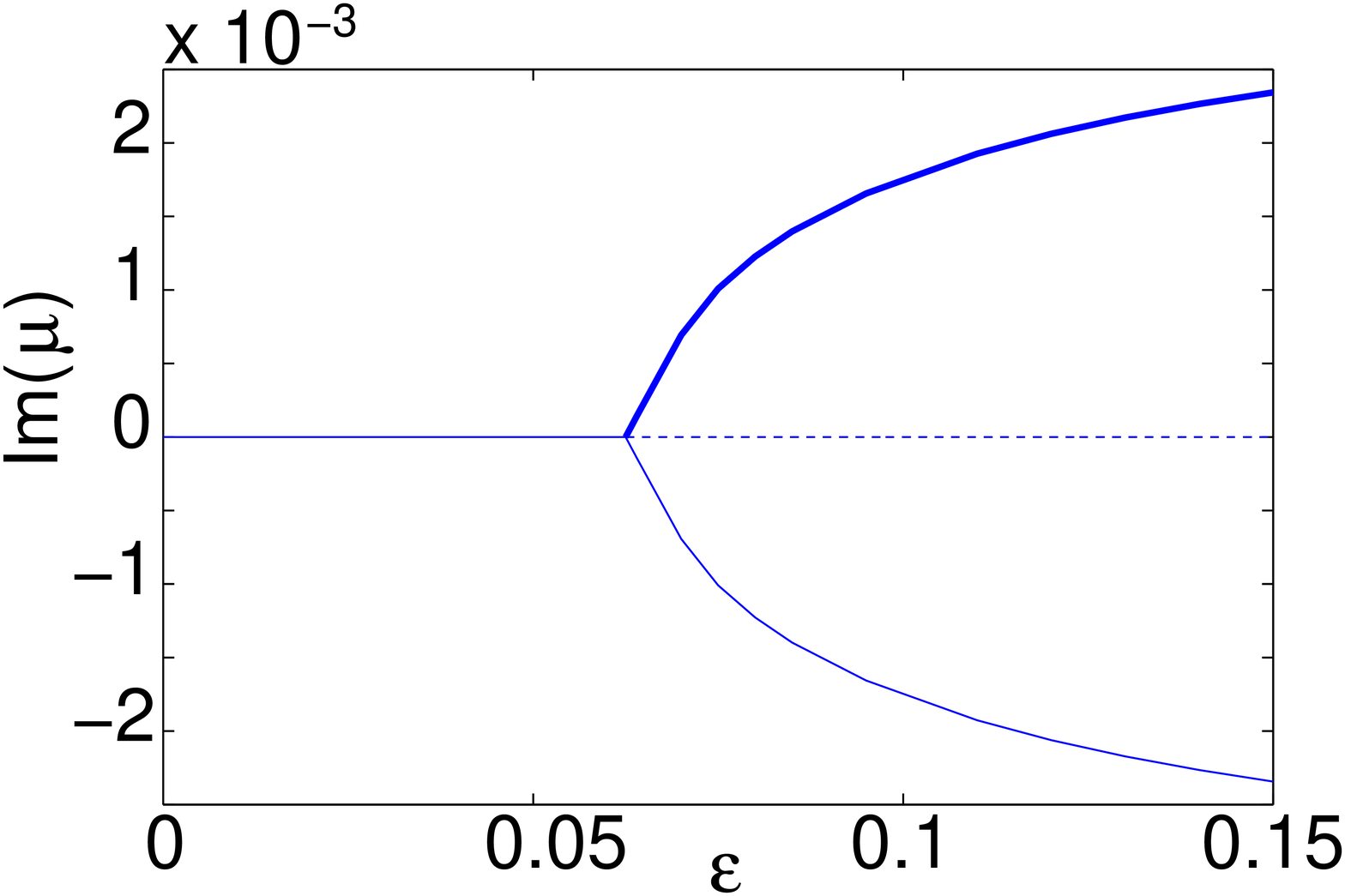}
\includegraphics[width=4cm]{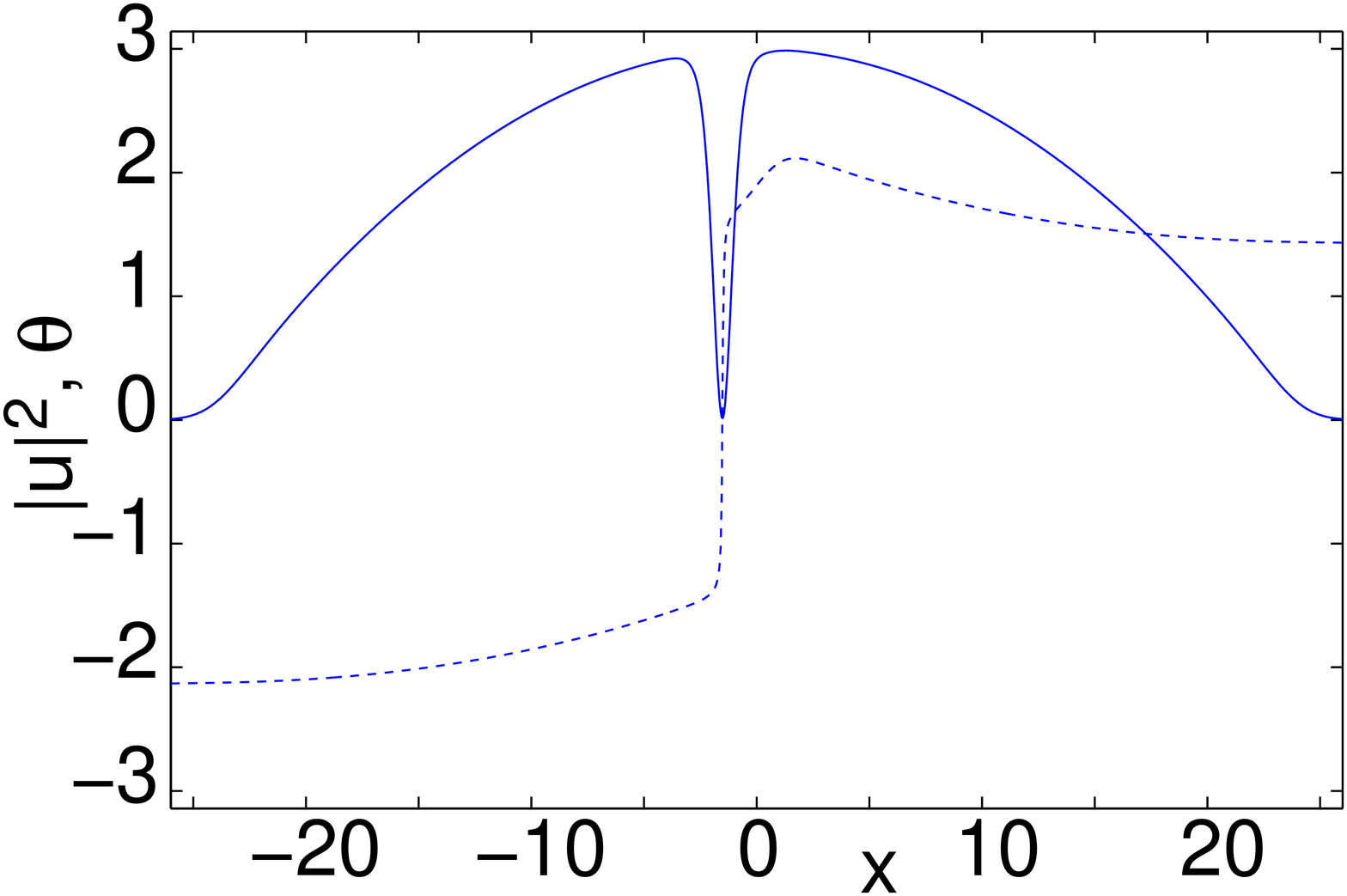}
~\includegraphics[width=4cm]{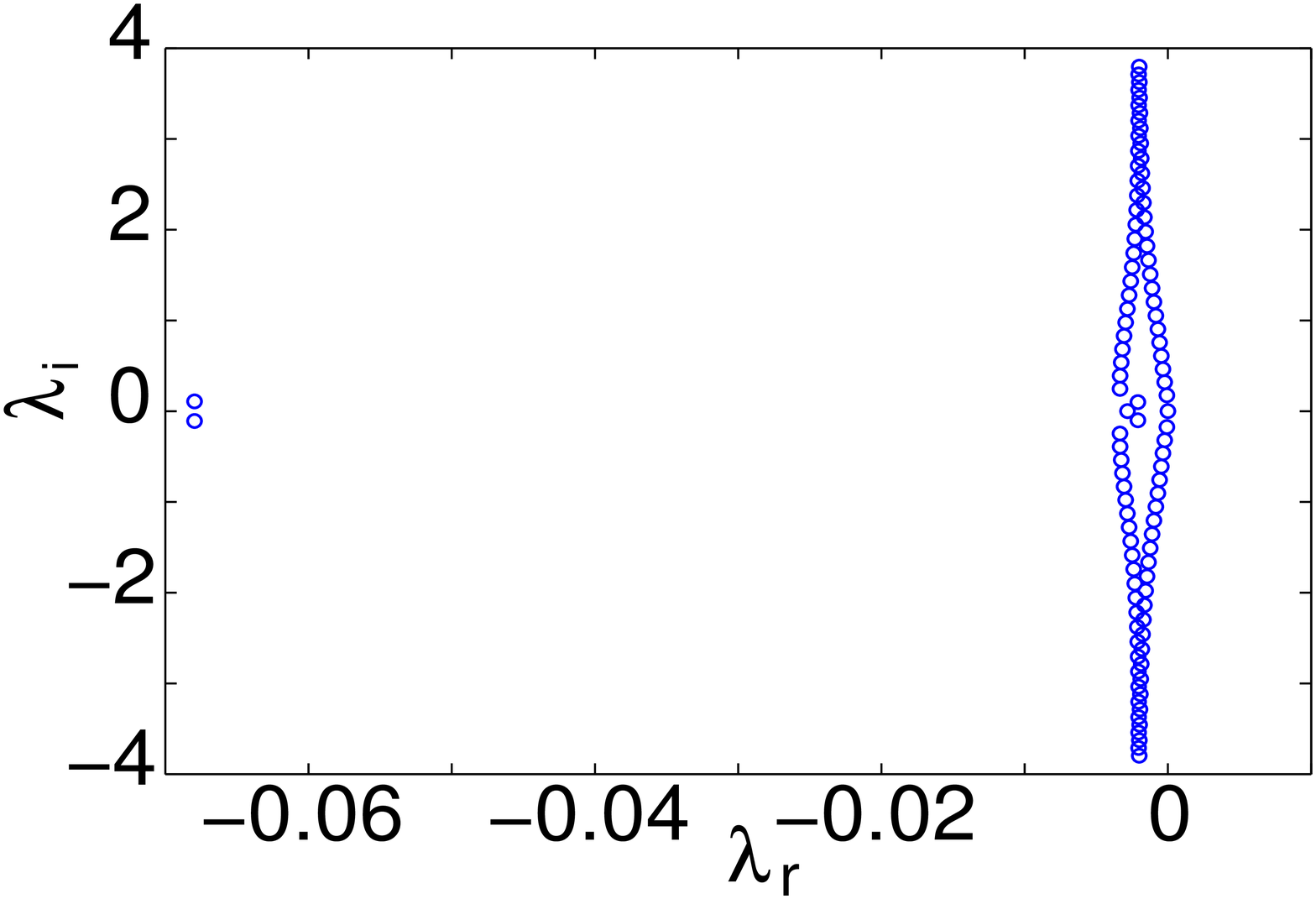}
\caption{(Color online) Properties of the ghost states.
The top left panel illustrates their (off-center) location with the
bold line illustrating the dynamically robust one. The center line corresponds
to the dark soliton at $x_0=0$ and the solid to dashed transition corresponds
to the SSB bifurcation. Another diagram illustrating the pitchfork character
of the transition is given in the top right panel through the
imaginary part of the chemical potential $\mu_I$ as a function of the
gain/loss parameter. Again, the persisting state is given in bold.
A typical example of the squared density and phase (solid and dashed,
respectively) of a ghost state for $\varepsilon=0.15$ is given in the
bottom left panel. Finally, a suggestive linear stability analysis
(as if this were a true steady state of the system) is performed
on the bottom right suggesting also the dynamical robustness of the relevant
state due to the absence of eigenvalues with a positive real part (i.e., of
ones corresponding to growth).}
\label{ghost1}
\end{figure}

The natural next questions then involve whether these states
can, in fact, be identified as such (exact solutions of the steady state
problem) and, perhaps more importantly, what is their role in the
observed dynamics. The first of these questions is answered in
Fig.~\ref{ghost1}, where we have performed a continuation of
these states as a function of $\varepsilon$, starting from the
Hamiltonian limit in the absence of gain/loss. The top left panel
of the figure presents the center of the solution as a function of
the parameter, while the top right one evaluates the imaginary
part of the chemical potential, by self-consistently enforcing
the condition~(\ref{chpot3}). It is key to mention here that for the
solutions with negative $x_0$, it is straightforward to see that they
correspond to a setting with $\mu_I >0$ [from Eq.~(\ref{chpot3})].
On the other hand, the term $e^{-i \mu t}$ will then involve
a part associated with $e^{-i (i \mu_I) t}$ and hence this state
will be associated with growth over time. On the other hand, the state
with   $\mu_I < 0$ will be connected to decay over time and does
not appear to be relevant for physical observations. It is for
that reason that one of the two states (the robust one) is denoted
with a bold solid line in the top panels of Fig.~\ref{ghost1}, while
the other one is denoted by a thin line (and unstable states centered
at $x_0=0$ in the diagram
are denoted by a dashed line).

First of all, this observation restores the canonical (and expected)
pitchfork nature of the observed bifurcation (see top panels
of Fig.~\ref{ghost1} and the daughter state of the bottom left panel).
But, at the same time it reconciles that nature with the absence
of any ``regular'' daughter states. A stability analysis of these
states is not particularly meaningful per se (as these are not stationary
states of the original problem),
yet it is natural to expect that it will be suggestive regarding
the evolution of perturbations in the vicinity of such states. This,
as shown in the bottom right panel for the displaced state of the bottom
left one in Fig.~\ref{ghost1}, illustrates the effective
stability of the resulting ghost state.

\begin{figure}[t]
\includegraphics[width=4cm]{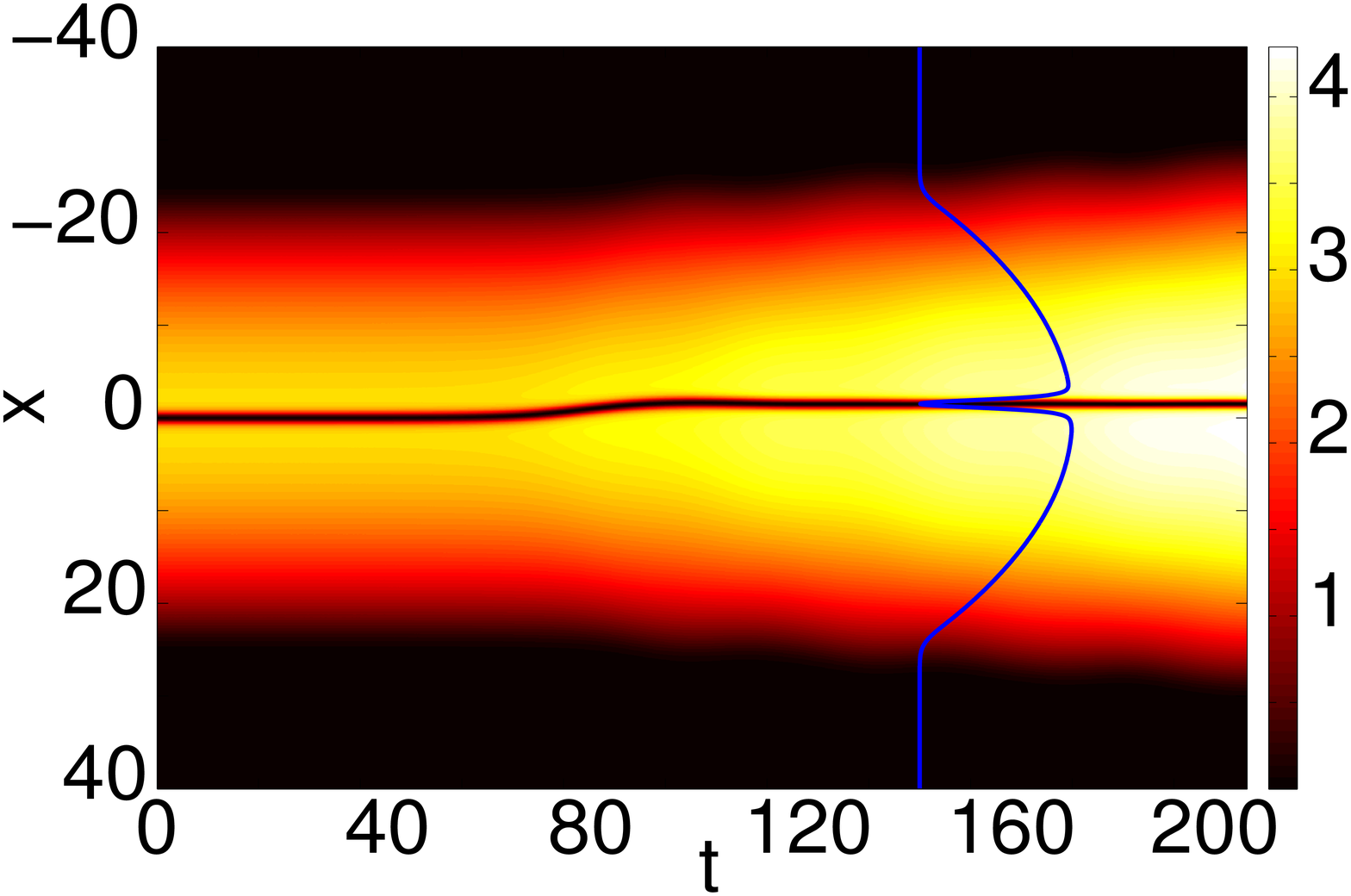}
~\includegraphics[width=4cm]{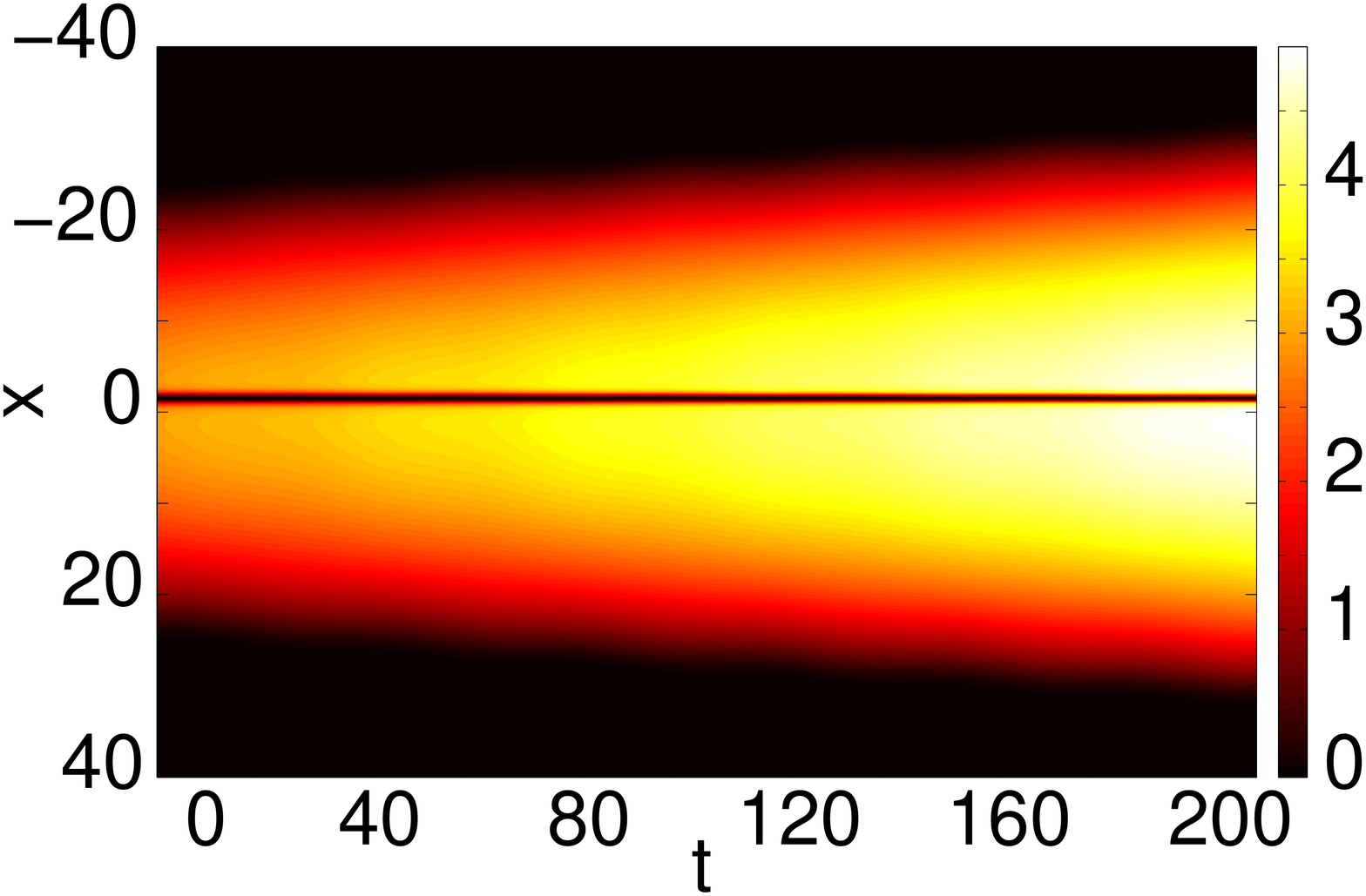}
\includegraphics[width=4cm]{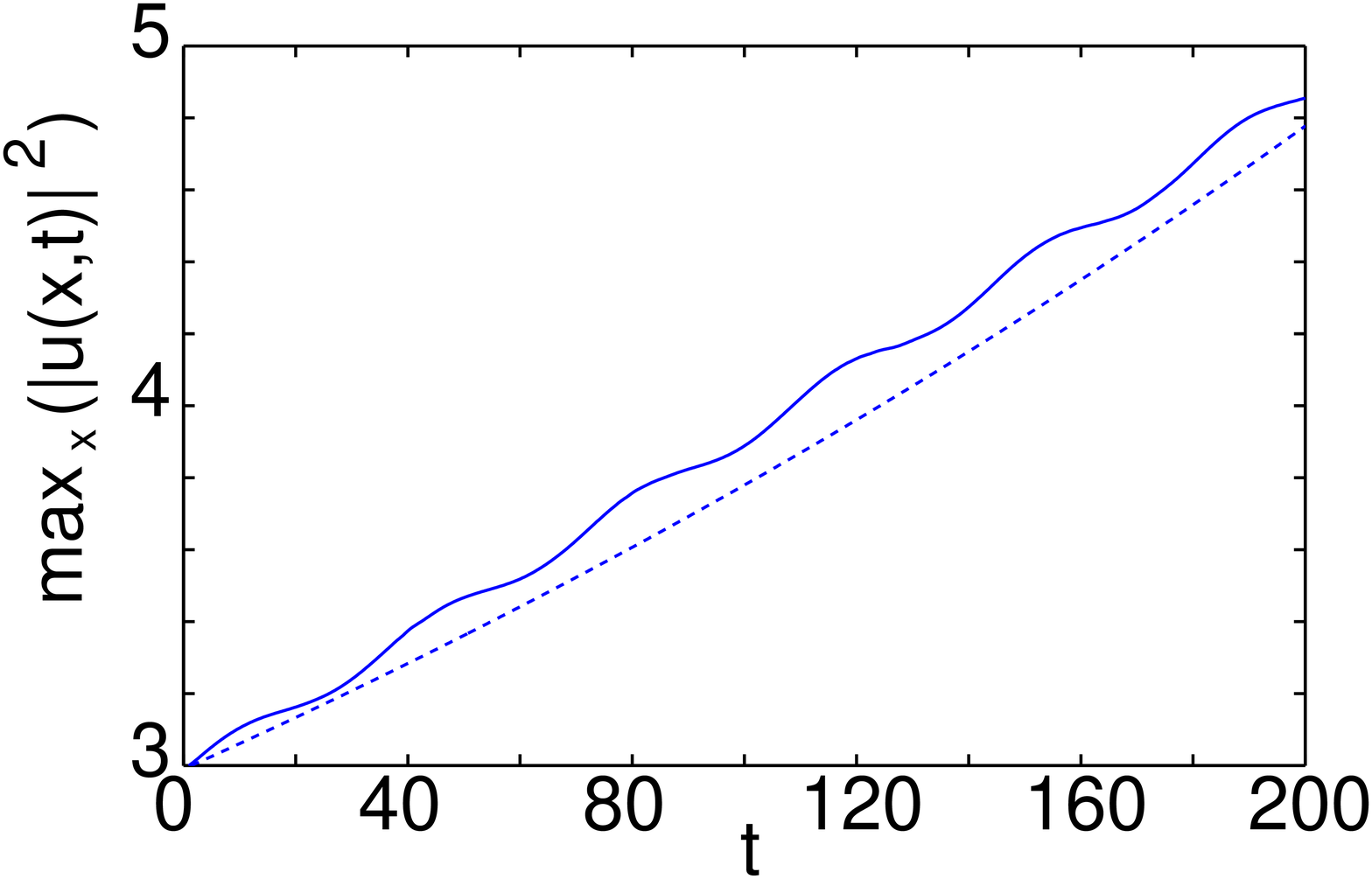}
\caption{(Color online) The top left panel shows the space-time evolution
of the contour plot of the unstable dark soliton at $x_0=0$
for $\varepsilon=0.15$. For comparison (and out of scale), we plot
(using a solid [blue] line) the
exact obtained ghost state for the same gain/loss parameter. Clearly the
dynamics gets attracted to this state, as is also corroborated by its
evolution as initial condition in the top right panel (the dark soliton
stays put, but the background width and amplitude grows in a way directly
analogous to the case on the left). Finally, the comparison of the
amplitude background growth $A(t)$ (solid line) with the law implied by
the growth rate of the ghost state $A(t)=A(0) \exp(\mu_I t)$ (dashed)
is shown on the bottom panel, suggesting good agreement between the two.}
\label{ghost2}
\end{figure}

We now turn to the dynamical implications of the existence of such
ghost modes. The asymmetry between the growth of the mode with the
dark soliton at $x_0<0$ and the decay of the one with $x_0>0$
justifies the observed asymmetry of the dynamical evolution of
the perturbations in Fig.~\ref{fig3}. To illustrate the relevance
of the ghost states in the
case where the random perturbation manifests
the instability of the $x_0=0$ soliton by kicking it to the left, we
performed some relevant numerical experiments in Fig.~\ref{ghost2},
for $\varepsilon=0.15$. Specifically, by initiating the soliton
at $x_0=0$, we confirm that it goes and sits precisely at the location
(for that particular $\varepsilon$) where the ghost state has the soliton
centered (cf.~with the out of scale profile of the shifted soliton
within the graph). On the other hand, for a clear comparison, in the right
panel of Fig.~\ref{ghost2}, we have
evolved the dynamical equations with exactly that proper
ghost state for the particular $\varepsilon$ as initial condition (wherein
the soliton is located at $-1.5$ for $\varepsilon=0.15$; see the right
panel of the figure). The dynamics clearly illustrates that the soliton
stays immobile (as expected by the nature of the ghost state), while
the background increases in amplitude and also the width of the solution
increases. These latter growth features are strongly reminiscent of
the evolution of the left panel of the figure past the point of destabilization
and convergence to a solitary wave centered at $x_0=-1.5$.
To further cement this proximity, we have computed the growth of the
background amplitude i.e., the maximum absolute value of the field $A(t)$
and have directly compared it to the simplistic law (implied by the
ghost state growth rate $\mu_I$) $A(t)=A(0) \exp(\mu_I t)$, yielding
the kind of agreement shown in the bottom panel of the figure.
The natural
conclusion is that while the ghost states are not genuine (steady)
attractors of the original dynamics, nevertheless the evolution of
the system closely shadows the relevant states and utilizes their
asymmetry.

As for the case where the dark soliton moves to $x_0>0$ due to the
perturbation, then the ghost state is simply untenable as a dynamical
state (due to its decay in time), hence the dark soliton proceeds to
escape from the region of influence of the gain and moves towards
a turning point. This motion imparts sufficient momentum to the solitary
wave that it can subsequently execute oscillations around $x_0=0$,
as shown in Fig.~\ref{fig3} (top right). Moreover, dynamical features
such as the generation of multi-solitons sitting on the lossy side
can be explained upon the consideration of ghost states associated with
multi-soliton states (see e.g.~the 3-soliton state of the bottom right
of Fig.~\ref{fig3}), while features such as those of the top panel
of Fig.~\ref{fig33} can be explained as well. The latter
can be comprehended on the basis of
a superposition of a ghost state centered on the lossy side
with a growing background, and of one (as in top left, or more
as in the top right panel of Fig.~\ref{ghost2}) soliton(s) moving to the opposite
direction and hence executing oscillations due to the absence of
a ghost state to trap it. The above characteristics of the ghost
states thus naturally lend themselves to the explanation of the
static and dynamic phenomenology observed herein.

\section{Symmetry Breaking and Nonlinear $\mathcal{PT}$
Phase Transitions in Two Dimensions.}
\label{sec:2D}

\begin{figure}[tbp]
\includegraphics[width=6.77cm]{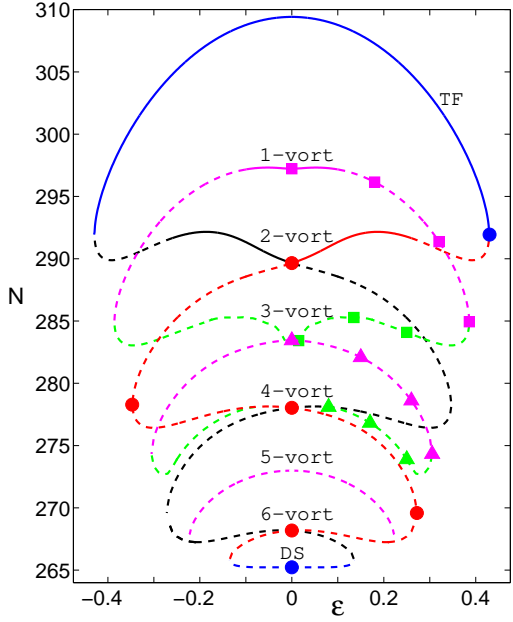}
\caption{(Color online)
Bifurcation diagram of stationary profiles in the 2D case.
Stable  and unstable branches, as determined by the corresponding BdG
analysis, are depicted, respectively, with solid and dashed lines.
The large symbols (circles, squares and triangles) denote the
location in parameter space for the series of profiles
depicted in Fig.~\ref{fig2D2}.
The parameter values are: $\mu=2$ and $\Omega=0.2$.
}
\label{fig2D1}
\end{figure}

\begin{figure}[tbp]
\includegraphics[width=2.82cm]{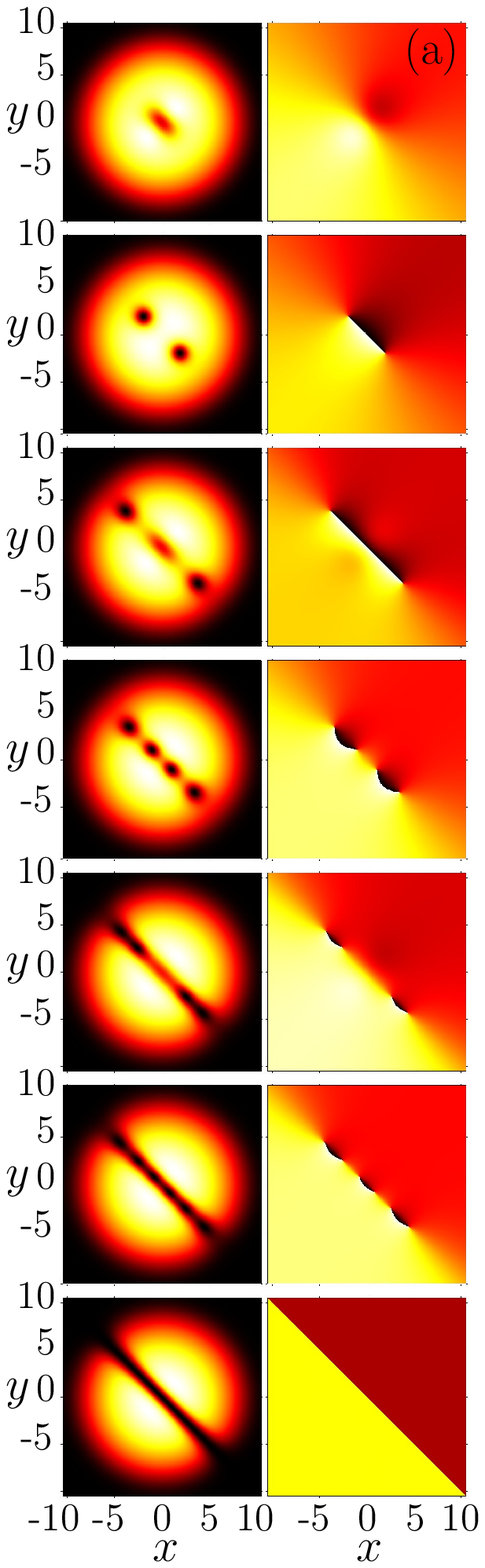}
\includegraphics[width=2.82cm]{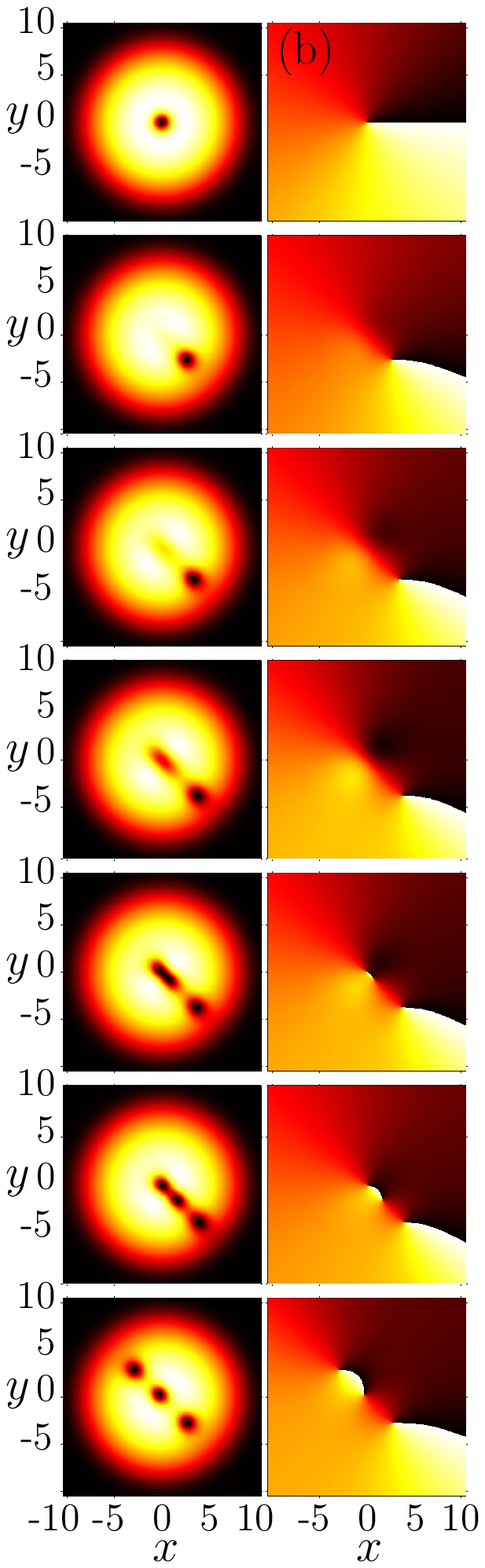}
\includegraphics[width=2.82cm]{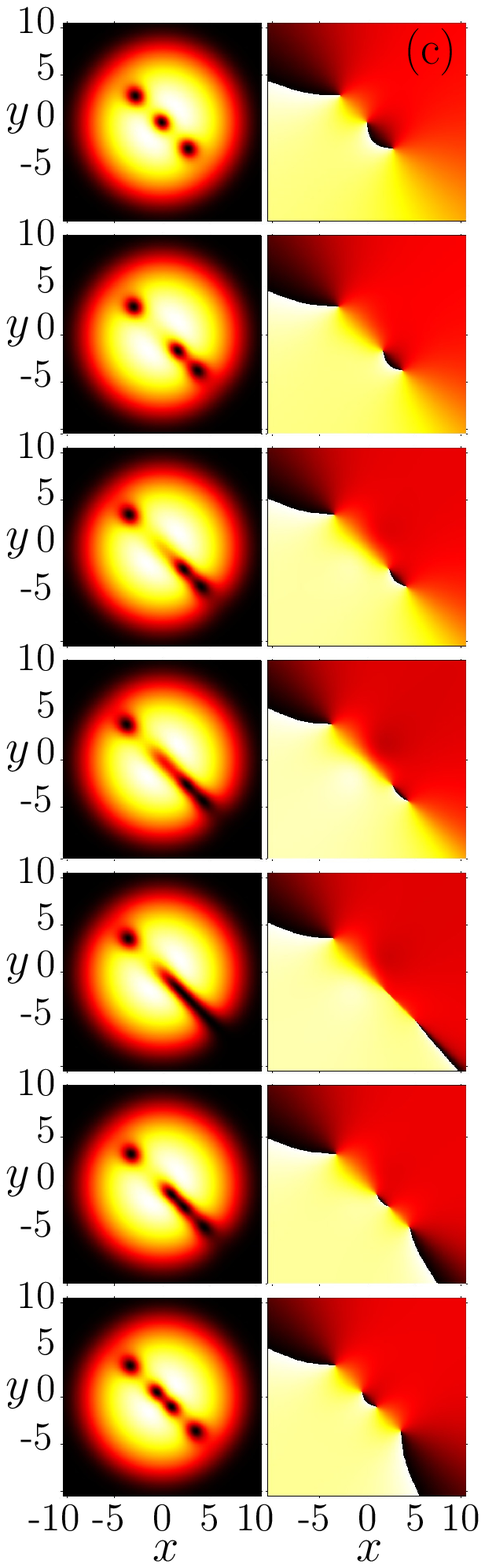}
\caption{(Color online)
Stationary states in the 2D case corresponding to the main
bifurcation branches in Fig.~\ref{fig2D1}.
The respective left (right) columns depict the density (phase)
profiles of the solutions.
(a) Profiles for the branch containing only even number of vortices
corresponding to the circles in Fig.~\ref{fig2D1} [from top to bottom].
(b) Profiles for the branch starting with a single vortex and connecting
with three vortex branch corresponding to the squares in Fig.~\ref{fig2D1}
[from top to bottom].
(c) Profiles for the branch starting with three symmetric
vortices and ending with four vortices corresponding to the triangles
in Fig.~\ref{fig2D1} [from top to bottom].
The parameter values are: $\mu=2$ and $\Omega=0.2$.}
\label{fig2D2}
\end{figure}

In order to consider the generalization of the ideas presented
herein to higher dimensions,
we now consider the case of a 2D $\mathcal{PT}$-symmetric potential
[$W(-x,-y)=-W(x,y)$] with
\begin{equation}
W(x,y)=\varepsilon (x+y) e^{-(x^2+y^2)/4}.
\label{W2D}
\end{equation}
The configurations sustained in the 2D setting follow a similar
symmetry breaking bifurcation scenario as in their 1D counterparts.
However, the bifurcation structure is naturally expected to be
more complex for the higher dimensional case.
Figure \ref{fig2D1} displays the bifurcation scenario for 2D
solutions with and without topological charge for $\Omega=0.2$.
The bifurcation diagram
includes the chargeless ground state (the TF background cloud),
solutions bearing from one to six vortices,
and the dark soliton stripe.
Similar to the 1D case, the TF background is stable in all its domain of
existence and collides, as $\varepsilon$ increases, with an excited state
(the two-vortex solution or vortex-dipole) in a blue-sky bifurcation
at $\varepsilon=\varepsilon_{\rm cr}^{(2)}$.
However, in contrast to the 1D case where this
collision occurs with the first excited state, in the 2D case the
collision happens with the {\em second} excited state consisting
of a vortex-dipole (a vortex pair with opposite charges).
This is due to the fact that the TF background has no
topological charge and, in turn, the vortex-dipole has also no
net topological charge. This allows for the emergence of the
vortex-dipole from a central dip on the TF background
[see top panel of Fig.~\ref{fig2D2}(a)] that becomes
deeper as $\varepsilon$ increases.
This blue-sky bifurcation is depicted in Fig.~\ref{fig2D1} where
the stable (center)
TF branch (solid blue curve denoted by {\tt TF}) collides with
the unstable (saddle)
vortex-dipole branch (red dashed curve denoted by {\tt 2-vort})
at the 1D equivalent of the critical point $\varepsilon=\varepsilon_{\rm cr}^{(2)}$.

It is possible to further follow the vortex-dipole branch by decreasing
$\varepsilon$ where we note that it regains stability for sufficiently
smaller values of $\varepsilon$ (see transition between the dashed and solid
portions of the red line for the branch denoted by {\tt 2-vort}).
As this branch is followed further (from top to bottom in Fig.~\ref{fig2D1})
a series of bifurcations arise where the vortices
are drawn towards the edge of the cloud, a central dip
deepens leading eventually to the emergence of a new vortex dipole
in the middle of the cloud (namely, the emergence of a
higher excited state).
Therefore, the branches with {\em even} number of vortices
are all inter-connected in this bifurcation scenario
and only the TF branch and part of the vortex-dipole branch are stable.
However, as more and more vortex pairs emerge, the extent of the
cloud ``saturates'' as it can no longer support new vortex pairs
and therefore is replaced by a dark soliton stripe (see dashed
blue line denoted by {\tt DS} in Fig.~\ref{fig2D1}).
This overall bifurcating scenario for even number of vortices
is symmetric such that if  $\varepsilon \rightarrow -\varepsilon$
the solutions are just flipped by $(x,y) \rightarrow (-x,-y)$.
Figure~\ref{fig2D2}(a) depicts the density and phase profiles for
different representatives along this branch.

Let us now describe the bifurcation scenario for {\em odd} number of
vortices. The first excited state, corresponding to a single vortex,
starts with a single vortex at the origin for $\varepsilon=0$ and it
is stable for small values of $\varepsilon$. As in the 1D case,
this first excited state suffers a SSB bifurcation at
$\varepsilon=\varepsilon_{\rm cr}^{(1)}$ where it loses its stability.
%
%
As $\varepsilon$ is increased, the single vortex moves towards
the periphery of the cloud and a dip at the center of the cloud
deepens until a vortex dipole emerges at the center in the same way
as new vortex-dipoles emerged for the branches with even number of
vortices described above.
This scenario connects, again through a blue-sky bifurcation,
the one-vortex branch (magenta line denoted
by {\tt 1-vort} in Fig.~\ref{fig2D1}) with the {\em asymmetric} three-vortex
($+ - +$ vortex tripole) branch (green dashed line denoted
by {\tt 3-vort} in Fig.~\ref{fig2D1}). The series of snapshots at the
parameters depicted by the squares in Fig.~\ref{fig2D1} is depicted in
Fig.~\ref{fig2D2}(b). It is nevertheless relevant
to note that the blue-sky bifurcation for the single-triple
vortex state (and more generally for any higher order pair
of vortices) happens for values of $\varepsilon$ lower than that
of the bifurcation involving the TF ground state and the vortex-dipole state.
Interestingly, although the structure of the relevant bifurcations
is reminiscent of its one-dimensional analogue,
there are also non-trivial differences, including the topological
charge issue
mentioned above, as well as
the order
in which the bifurcations occur (between more fundamental and more highly
excited states) which is reversed.
As it is evident from the figure,
the asymmetric three-vortex branch eventually connects with the
symmetric one for values of $\varepsilon \rightarrow 0$.
A similar bifurcation occurs with the symmetric three-vortex
branch, which becomes asymmetric with a deepening dip at the
center where a vortex pair emerges (at the same time that a
vortex is lost at the periphery),
connecting in this way with the
four-vortex branch [see series of snapshots in panels \ref{fig2D2}(c)].
As a relevant aside, let us also mention that all of the above
true stationary states of the system (with real chemical potential
or propagation constant) clearly form along the diagonal $x+y=0$
on which there is no gain or loss according to the prescription
of Eq.~(\ref{W2D}).

\begin{figure}[tbp]
\includegraphics[width=1.95cm]{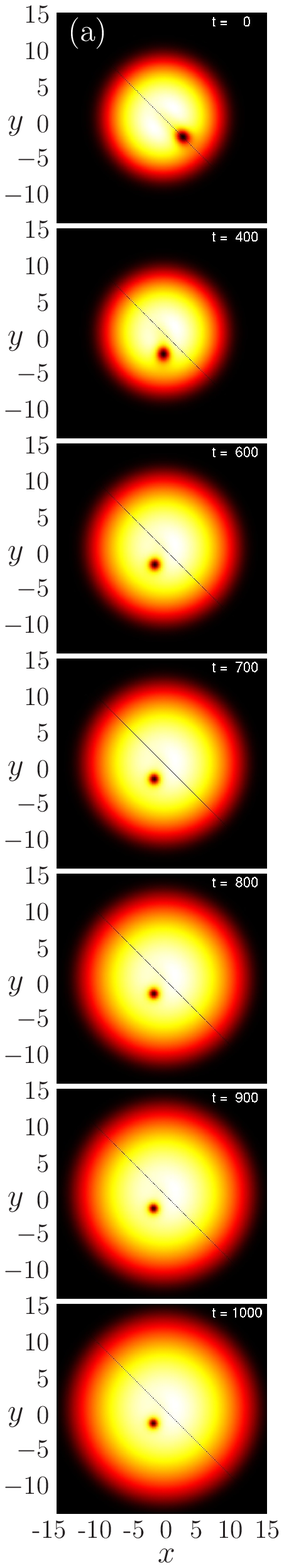}
\includegraphics[width=1.95cm]{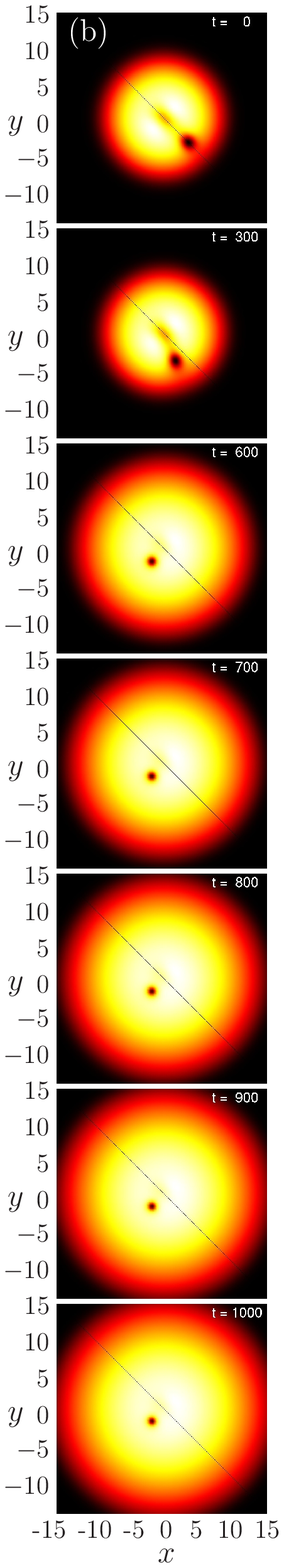}
\includegraphics[width=1.95cm]{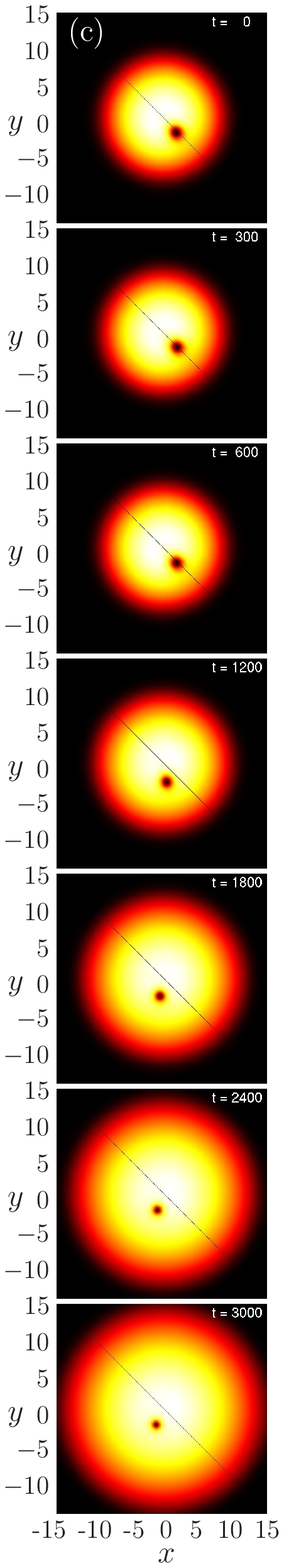}
\includegraphics[width=1.95cm]{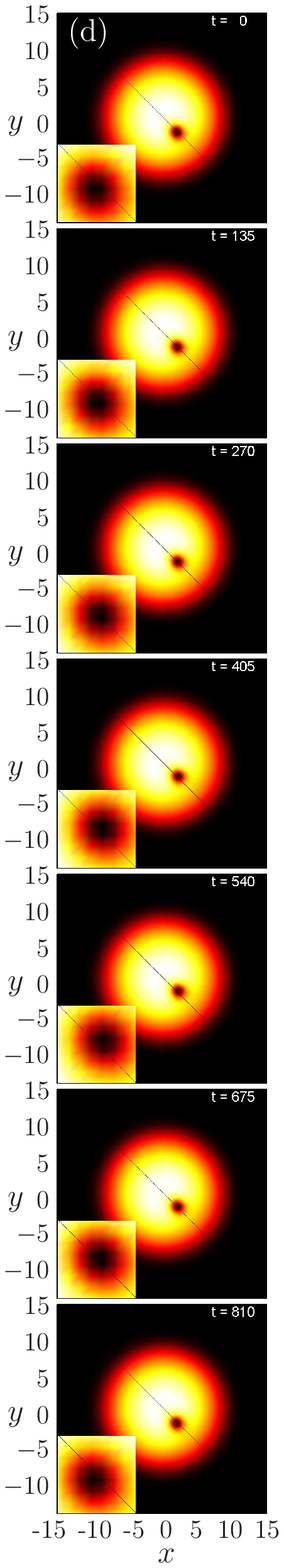}
\caption{(Color online)
Dynamics of unstable steady states in the 2D case.
All panels depict the density profiles at the times indicated
for $\mu=2$ and $\Omega=0.2$.
The thin diagonal line in all panels depicts the border
between the loss and gain sides of $W(x,y)$.
(a-b) Unstable single vortex state for $\varepsilon=0.2$
and $\varepsilon=0.35$, respectively.
(c) Stable single vortex configuration for $\varepsilon=0.1$
perturbed by a small rotation of 5 degrees.
(d) Stable single vortex configuration for $\varepsilon=0.1$
perturbed by a small rotation of 4 degrees performs a
periodic oscillation around its original position (a single
period is shown, see zoom in the inset).
}
\label{fig2D_dyn1}
\end{figure}

It is important to mention that the precise structure of the bifurcation
diagram depends on the values of the trap strength $\Omega$ and the
propagation constant $\mu$. In general terms, for weaker $\Omega$ and/or
larger $\mu$ the spatial extent of the TF background will be larger
and thus allowing for a longer bifurcating chain of higher-order
vortex states (before all the vortices finally merge into a dark
soliton stripe).
%
%
Nonetheless, the displayed SSB instabilities and the
nonlinear $\mathcal{PT}$ phase
transition involving the cascade of blue-sky
bifurcations appear to be universal in confining $\mathcal{PT}$-symmetric
potentials as considered herein.

\begin{figure}[tbp]
\includegraphics[width=1.95cm]{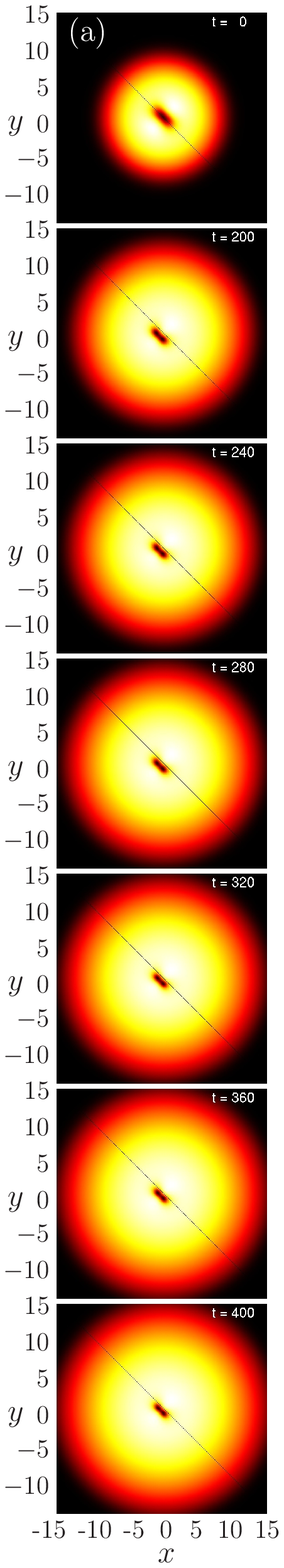}
\includegraphics[width=1.95cm]{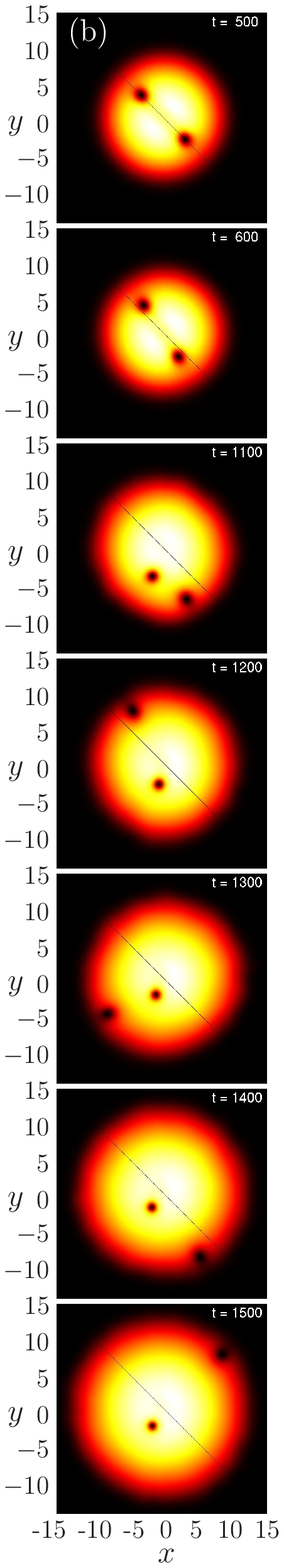}
\caption{(Color online)
Same as in Fig.~\ref{fig2D_dyn1} for the following
two-vortex scenarios.
(a) Unstable tight vortex-dipole state belonging to the
branch directly connecting with the TF cloud (depicted
by the red dashed line denoted by {\tt 2-vort} in Fig.~\ref{fig2D1})
for $\varepsilon=0.4$.
(b) Unstable, well separated, two-vortex state belonging to the
branch connecting with the four-vortex branch (depicted
by the black dashed line denoted by {\tt 2-vort} in Fig.~\ref{fig2D1})
for $\varepsilon=0.2$.
}
\label{fig2D_dyn2}
\end{figure}

Let us now describe the dynamics of the unstable steady states
described in the above bifurcation scenario.
We first start by describing the unstable dynamics of single
vortex states.
In this case, in analogy with the 1D case, single vortex states
tend to migrate towards the lossy side of the potential.
This tendency is depicted by the series of density snapshots in
Figs.~\ref{fig2D_dyn1}(a) and \ref{fig2D_dyn1}(b) where {\em unstable}
single vortex states for $\varepsilon=0.2$ and $\varepsilon=0.35$, respectively,
migrate to different positions within the lossy side. This phenomenology
can be very clearly understood on the basis of our ghost state interpretation.
Once again, ghost states are present here corresponding to a unique
shift of the vortex center to a location such that $x_0+y_0 < 0$.
This state is associated with $\mu_I>0$, due to the fact that
$\int W(x,y) |u(x,y)|^2 dx dy > 0$ and hence leads to growth of the
amplitude and width of the condensate, a trait which can be clearly
observed in the relevant panels.
It is also important to mention that the stable single vortex, close
to where it loses stability, is only weakly stable and can be `kicked out'
by a relatively small perturbation towards the lossy side.
This scenario is depicted in
Fig.~\ref{fig2D_dyn1}(c) where the {\em stable} single vortex
configuration for $\varepsilon=0.1$ is perturbed by applying a small
rotation of 5 degrees. Although the rotation is very small and
that the unperturbed configuration is stable, we
note that the vortex eventually migrates (after a long transient, see that
time in the series of snapshots runs to t=3000) towards the
lossy side. Further numerics shows that angles larger than 4 degrees
display the same dynamics while smaller angles correspond to
the single vortex returning periodically towards its original position
[see for example the periodic orbit depicted in Fig.~\ref{fig2D_dyn1}(d)
where the stable vortex is initially rotated by 4 degrees].
%

\begin{figure}[tbp]
\includegraphics[width=1.95cm]{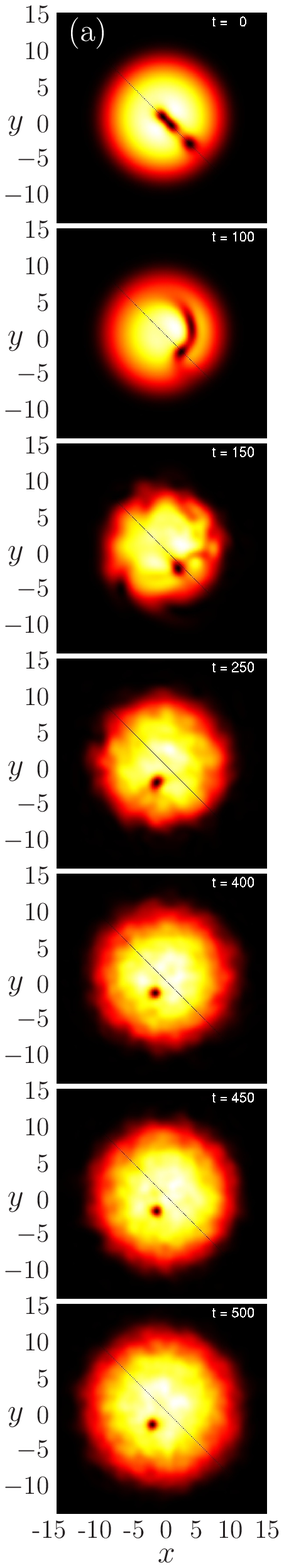}
\includegraphics[width=1.95cm]{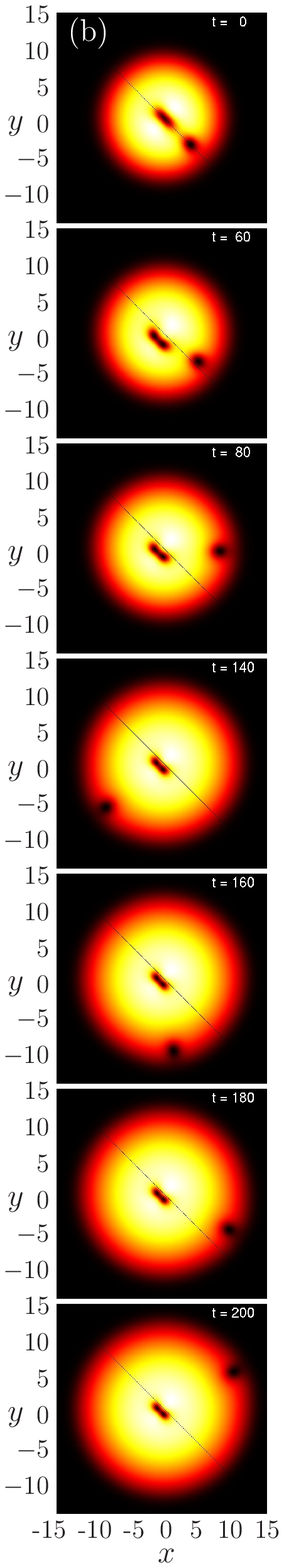}
\includegraphics[width=1.95cm]{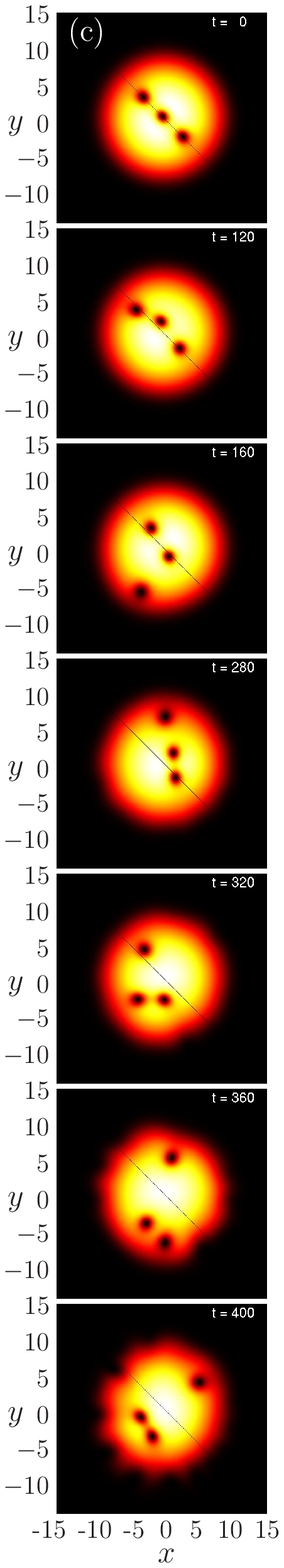}
\includegraphics[width=1.95cm]{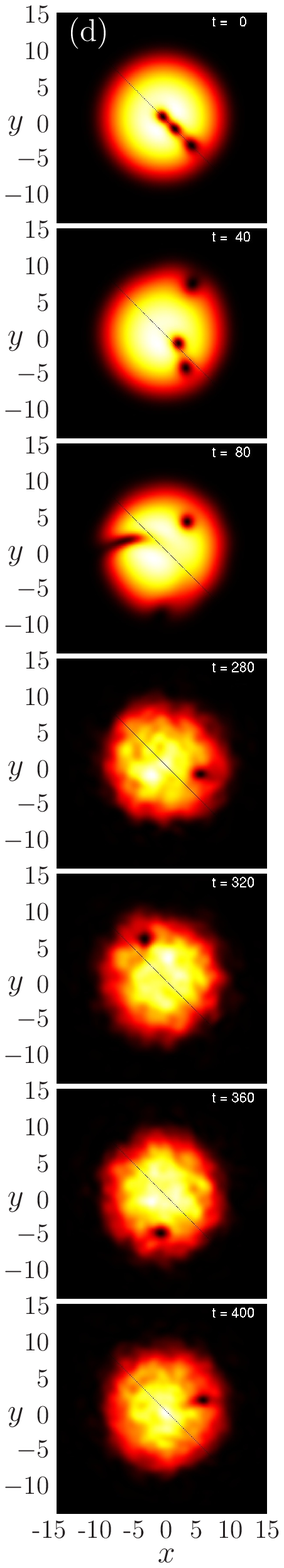}
\caption{(Color online)
Same as in Fig.~\ref{fig2D_dyn1} for the following
three-vortex scenarios.
(a-b) Unstable asymmetric three-vortex states (depicted by the
green dashed line denoted by {\tt 3-vort} in Fig.~\ref{fig2D1})
for $\varepsilon=0.2$ and $\varepsilon=0.35$, respectively.
(c) Unstable symmetric three-vortex state
for $\varepsilon=0$.
(d) Unstable asymmetric three-vortex state (depicted by the
green dashed line denoted by {\tt 3-vort} in Fig.~\ref{fig2D1})
for $\varepsilon=0.1$.
}
\label{fig2D_dyn3}
\end{figure}

Let us now describe the evolution of unstable vortex-dipoles.
In Fig.~\ref{fig2D_dyn2}(a) we depict the evolution of a {\em tight}
vortex-dipole close to the parameter values where it is created
(i.e., in the dip at the center of the cloud). As the figure shows,
the tight vortex-dipole
undergoes a small excursion into the lossy side and, as time progresses,
it re-approaches
to the center of the cloud at the same time that the cloud
grows (as in the 1D case). Clearly the relevant interpretation
here involves the existence of a ghost state which attracts the
dynamics and contains both vortex members of the dipole on the lossy
side of the system.
In Fig.~\ref{fig2D_dyn2}(b) we depict the dynamics arising from a
{\em well separated} unstable vortex-dipole state. We typically observe
that, for a well separated unstable vortex dipole, one vortex is
trapped at a suitable location within
the lossy side while the other vortex
circles around slowly approaching the periphery of the cloud where
it is finally absorbed. This is in a number of ways tantamount to
the two-soliton dynamics that we explored in Fig.~\ref{fig33}.
Generally, it is clear that the 3 possible dynamical scenaria
for the vortices constituting the dipole and involving
ghost states here are that either both vortices transition to the
loss side [Fig.~\ref{fig2D_dyn2}(a)], or one [Fig.~\ref{fig2D_dyn2}(b)]
or none (not shown here).

\begin{figure}[tbp]
\includegraphics[width=1.95cm]{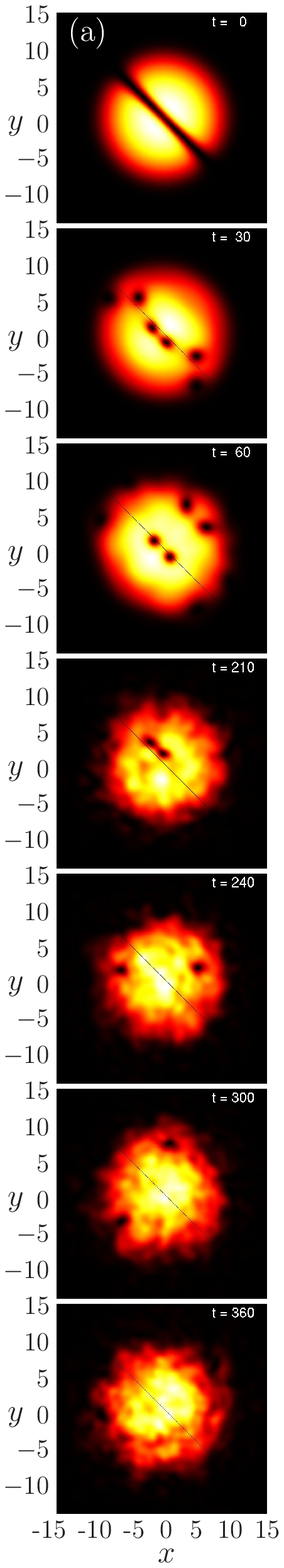}
\includegraphics[width=1.95cm]{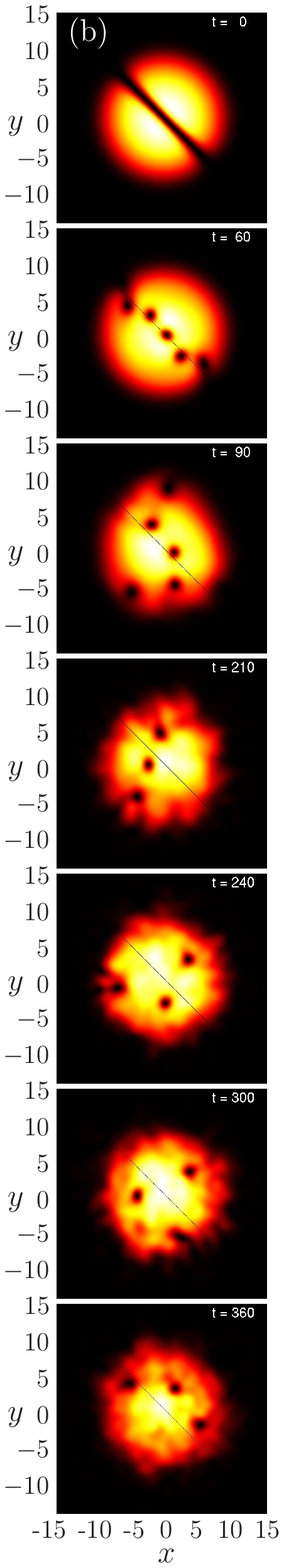}
\includegraphics[width=1.95cm]{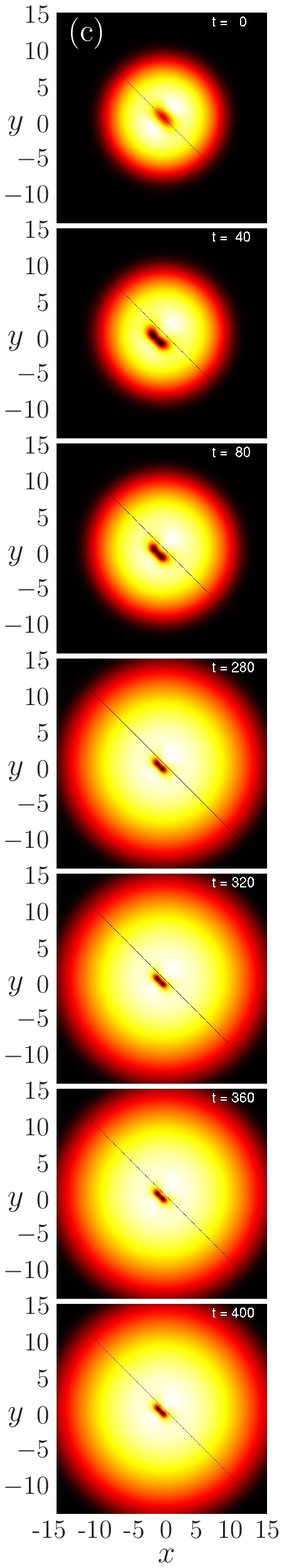}
\caption{(Color online)
Same as in Fig.~\ref{fig2D_dyn1} for the following scenarios.
(a-b) Unstable dark soliton stripes (depicted by the
lowest blue dashed line denoted by {\tt DS} in Fig.~\ref{fig2D1})
for $\varepsilon=0.1$ and $\varepsilon=0$, respectively.
(c) Thomas-Fermi (ground) state obtained for $\varepsilon=0.4$ and let to
evolve for $\varepsilon=0.43$.
}
\label{fig2D_dyn4}
\end{figure}

Let us now briefly describe some typical evolution examples
for higher order unstable states. In particular,
let us start by discussing the three-vortex states.
Figure~\ref{fig2D_dyn3}(a) depicts the evolution of an unstable asymmetric
3-vortex profile. In this case, the outer vortex migrates towards the
attracting basin within the lossy side while the remaining
vortex-dipole gets violently ejected and the two vortices annhilate
each other. This again confirms the dynamical
relevance of the single-vortex corresponding ghost state.
Another possible scenario for the 3-vortex configuration is depicted
in Fig.~\ref{fig2D_dyn3}(b) where this time the central vortex-dipole
slowly migrates towards the lossy side and slowly re-approaches the
center of the trap as time progresses while the outer vortex performs
large excursions around the periphery of the cloud where it is
eventually absorbed. In this setting, we observe once again the dynamics
of a vortex-dipole ghost state (with the background in this, as well
as in the previous example featuring the analyzed growth and
concurrent spreading).
%
%
For contrasting purposes, we depict in Fig.~\ref{fig2D_dyn3}(c) the
dynamics of an unstable {\em symmetric} three-vortex state for
$\varepsilon=0$. In this case, the external potential is
conservative and the motion of the three vortices is
Hamiltonian and thus, typically, the three vortices will remain
orbiting inside the cloud for very long times. Nevertheless,
notice in the latter case the well established (and generic)
instability of
the Hamiltonian vortex tripole which leads to its break-up into a
vortex dipole and a single vortex~\cite{middelk1,stock1}.
Finally, in Fig.~\ref{fig2D_dyn3}(d) we depict the evolution of
another unstable asymmetric state. In this case it can be observed
that two vortices annhilate each other and the third does not
get absorbed by the lossy side as in Fig.~\ref{fig2D_dyn3}(a),
but it performs large oscillations close to the periphery of the
cloud where it is finally absorbed. The precession of this vortex
due to its initial placement within the gain side suggests a direct
analogy of this case (and after the dipole annihilation) with the
top left panel of Fig.~\ref{fig3}.
%

Finally, we briefly explore the dynamics of unstable dark soliton
stripes. Dark soliton stripes are known to be unstable
in the Hamiltonian case ($\varepsilon=0$) due to the
so-called snaking instability \cite{kuz2,kuz3} (although they can
be rendered stable by a very tightly confining
trap \cite{PanosAvoidingRedCatastrophe}
or a wall-like external potential \cite{Manjun}).
Figure \ref{fig2D_dyn4}(a) depicts the typical evolution of an
unstable dark soliton stripe solution in the non-Hamiltonian
($\mathcal{PT}$-symmetric) case. In this case, the
dark soliton stripe decays into a chain of alternating charge
vortices induced by the snaking instability and their
long term behavior includes annihilation
and large oscillations where they get eventually absorbed
at the periphery of the cloud leaving behind a highly perturbed
(chargeless) Thomas-Fermi ground state.
In contrast, as depicted in Fig.~\ref{fig2D_dyn4}(b), for the
Hamiltonian case of $\varepsilon=0$, most of the vortices that formed
from the snaking instability remain interacting for long times
(i.e., the latter are more likely to survive here, analogously
to what we also saw for the tripole case above).
Finally, in Fig.~\ref{fig2D_dyn4}(c) we depict the evolution when
starting with a TF cloud for $\varepsilon$ past the critical
value of existence of the stationary TF cloud. In this example,
we compute the stationary (stable), chargeless, steady TF profile
for $\varepsilon=0.4$ and run it with $\varepsilon=0.43$ (i.e.,
past the blue-sky bifurcation between the TF cloud and the vortex
dipole states). We observe in this simulation that the central
dip of the TF (which contains no charge) rapidly gets converted
into a tight vortex-dipole state which does a small excursion
towards the lossy side [very similar to the dynamics of the
unstable tight vortex-dipole profile of Fig.~\ref{fig2D_dyn2}(a)],
clearly suggesting the persistence and dynamical selection of the
corresponding vortex-dipole ghost state.
This behavior is akin to what we observed in 1D for values of
$\varepsilon$ past $\varepsilon_{\rm cr}^{(2)}$ (the dark soliton
`sprinkler' case) where excited states with topological
charge are dynamically created from chargeless configurations.
Hence, the two-dimensional setting may be used as a
``vortex sprinkler''.




\section{Conclusions.}

In the present work, we offered a basic, yet  self-contained
view on the fundamental modifications that the phenomenology of nonlinear
entities sustains in the presence of $\mathcal{PT}$-symmetric confining
potentials (although the case where there is no real potential was
also touched upon as well).
The prototypical among the observed features of such systems were
found to be the emergence of symmetry breaking bifurcations that
destabilizes single, triple, five, etc. soliton states in one
dimension and similarly one- or two- or higher vortex states
in two-dimensions.  The second important and unexpected feature
was the emergence of a nonlinear analogue of the
$\mathcal{PT}$-phase-transition, whereby e.g.~the ground and
first excited, the second and third excited, the fourth and fifth
excited state and so on (in 1D) are led to pairwise collisions and blue-sky
bifurcations.
Generalizations
of these effects which, however, properly respect the topological nature
of the states (e.g.~the zero-vortex with the vortex dipole, the single
vortex with the vortex tripole etc.) are found to arise in the two-dimensional
setting. In addition, the pitchfork nature of the symmetry-breaking
bifurcation was illustrated through the examination of the so-called
ghost states and the proof, not only of their genuinely complex
eigenvalue parameter, but also of their direct relevance for the
dynamics of the full problem, despite the fact that they do not constitute
exact solutions thereof (but rather only solutions of the static problem).
The dynamical evolution of unstable soliton and vortex states was
examined for all the identified instabilities, revealing a wealth
of possibilities involving ghost states, soliton/vortex sprinklers,
and oscillations between the loss and gain sides.

It would be interesting to offer a detailed map of the
ghost states of the system. Here, we have illustrated their
existence and observed many of their dynamical implications,
yet a systematic characterization of their emergence
(and perhaps the more challenging understanding of their
basins of attraction) certainly merits further examination.
On the other hand, a feature that has been examined in some
detail in the context of $\mathcal{PT}$-symmetric linear
problems which is, arguably, worthwhile to consider here
concerns the analytic continuation of the states past
the point of the $\mathcal{PT}$-phase-transition.
This was already touched upon in the nonlinear realm for
double well potentials in the work of Refs.~\cite{R34,R44,R45,R46}, but
it would be relevant to offer a definitive view of such
analytically continued states for general potentials.
Finally,
a canonical set of investigations which is still missing concerns the effects
of such potentials in three-dimensional continuum or one- and
higher-dimensional (infinite) lattice contexts.
These topics will be pursued in future publications.



\end{document}